\documentclass[12pt]{article}\usepackage[hyperfootnotes=false]{hyperref}
   \usepackage{epsfig}
      \usepackage{amsmath}
      \usepackage{amssymb}
  \usepackage{graphicx}
  \newlength{\abstractwidth}
  \renewcommand{\thefootnote}{\fnsymbol{footnote}}
  \renewcommand{\thanks}[1]{\footnote{#1}} 
  \newcommand{\starttext}{
  \setcounter{footnote}{0}
  \renewcommand{\thefootnote}{\arabic{footnote}}}
  \renewcommand{\theequation}{\thesection.\arabic{equation}}
  \newcommand{\be}{\begin{equation}}
  \newcommand{\bea}{\begin{eqnarray}}
  \newcommand{\eea}{\end{eqnarray}}
  \newcommand{\beq}{\begin{equation}}
  \newcommand{\ee}{\end{equation}}
  \newcommand{\eeq}{\end{equation}}

  \renewcommand{\a}{\alpha}
  \renewcommand{\b}{\beta}

  \def\ba{\begin{eqnarray}}
  \def\ea{\end{eqnarray}}

  \def\12{{1 \over 2}}
  \def\eq{&=&}

 \def\la{\langle}
  \def\ra{\rangle}
  
 \def\simleq{\; \raise0.3ex\hbox{$<$\kern-0.75em
      \raise-1.1ex\hbox{$\sim$}}\; }
 \def\simgeq{\; \raise0.3ex\hbox{$>$\kern-0.75em
      \raise-1.1ex\hbox{$\sim$}}\; }

\def\O2{\Omega_2}

\def\bi{\begin{itemize}}
  \def\ei{\end{itemize}}

\def\S{Schwarzschild}
\def\sc{\setcounter{equation}{0}}

\def\r{${\cal{R}}$}

\def\h{${\cal{H}}$}

\def\W{$\Omega$}
\def\W'{$\Omega$}

\def\V{\Omega}
\def\V'{\Omega}
\def\dof{degrees of freedom  }
\def\a{${\cal{A}}$}
\def\b{${\cal{B}}$}

\begin{document}
  \renewcommand{\theequation}{\thesection.\arabic{equation}}

\begin{titlepage}
  \rightline{}
  \bigskip

  \bigskip\bigskip\bigskip\bigskip

    \centerline{\Large \bf {Black Hole Complementarity and the }}
    \bigskip
\centerline{\Large \bf { Harlow-Hayden Conjecture$^{\dag}$ }}
    \bigskip

  \begin{center}
 \bf {{  Leonard Susskind}}
  \bigskip \rm

 Stanford Institute for Theoretical Physics and  Department of Physics, Stanford University\\
Stanford, CA 94305-4060, USA \\
\rm

\bigskip
\bigskip

  $^{\dag}$ \bf On the occasion of  John Preskill's 60th birthday. \rm

\vspace{2cm}
  \end{center}

  \bigskip\bigskip

 \bigskip\bigskip
  \begin{abstract}

Black hole complementarity, as originally  formulated in the 1990's by  Preskill,  't Hooft, and myself  is now being  challenged by the  Almheiri-Marolf-Polchinski-Sully firewall argument.
 The AMPS argument relies on an implicit assumption---the ``proximity postulate"---which says that the interior  of a black hole must be constructed from degrees of freedom that are physically near the black hole. The proximity postulate manifestly contradicts the idea that  interior information is redundant with information in Hawking radiation, which is very far from the black hole.  AMPS argue that a violation of the proximity postulate would lead to a contradiction in a thought-experiment in which Alice distills the Hawking radiation and brings a bit back to the black hole. According to AMPS  the only way to protect against the contradiction is for a firewall to form at the Page time. But the measurement that Alice must make, is of such a fine-grained nature that carrying it out before the black hole evaporates may be impossible.  Harlow and Hayden have found evidence that the limits of quantum computation do in fact prevent Alice from carrying out her experiment in less than exponential time.
 If their conjecture is correct then black hole complementarity may be alive and well.

 My aim here is to give an overview of the firewall argument, and its basis in  the proximity postulate; as well as the counterargument based on computational complexity, as conjectured by Harlow and Hayden.

 \medskip
  \noindent
  \end{abstract}

  \end{titlepage}

    \starttext \baselineskip=17.63pt \setcounter{footnote}{0}

  \tableofcontents

\sc
\section{Introductory Remarks}

\bigskip

The  claim of black hole complementarity (BHC) \cite{Susskind:1993if} is that information is not invariantly localized. Under certain conditions  a bit can appear to be  ``here" to one observer, and  far away to another. The ambiguous nature of localization was codified in BHC and the holographic principle. John Preskill, Gerard 't Hooft, and I championed this view in the early and mid 1990's. Since then it has become an accepted principle, particularly after gauge-gravity dualities were discovered.

Recently  BHC  has been  challenged by  Almheiri, Marolf, Polchinski and Sully  (AMPS) \cite{Almheiri:2012rt}. who  argued that the postulates of BHC are mutually inconsistent; in particular, they claim that  the purity postulate, the assumption of semiclassical QFT outside the black hole,  and the no-drama postulate lead to a contradiction\footnote{The firewall argument has been challenged or supported in a number of papers, making a variety of arguments which may be related to the arguments presented here. Here is a selection \cite{Nomura:2012sw}.}.

If the firewall argument is correct then it may represent a  step backward to a more traditional idea of information-localization, but at the cost of a   breakdown in the concept of a smooth horizon. At a recent firewall meeting at Stanford the confusion was so great that it prompted one famous black hole theorist to say that we are right back where we were forty years ago. John Preskill even suggested that we backtrack on the principle---by now a consensus belief---that information is conserved in black hole evaporation. Personally I found the AMPS argument compelling enough that I wrote a  paper entitled ``The Transfer of Entanglement: The Case for Firewalls" \cite{Susskind:2012uw}.

I can't prove it, but I think the firewall controversy may eventually turn out to be a false alarm, but one with an extremely interesting lesson.  This occasion  gives me an opportunity to discuss that lesson and to explain why BHC, as originally envisioned by Preskill, 't Hooft, and myself, may outlive the current controversy.

\subsection{Postulate 5}
When  asked if  BHC means that information behind the horizon is meaningless I've generally answered no; information in the interior of a black hole is meaningful, but  it is redundant with information in the exterior of the black hole.  At early time, before there's been much evaporation, the redundancy is between the interior and  the stretched horizon. Later, after a great deal of evaporation, the redundancy is between the interior and the Hawking radiation. The interior is meaningful, but  it, and  the Hawking radiation, should not be counted as independent.

One  implicit assumption of the AMPS paper explicitly contradicts the above statement of BHC. The authors assume, as in the original BHC paper,  that the \dof \ of the interior must be constructed from exterior \dof.  But they also assume that  those exterior \dof \  are physically near the black hole. In other words AMPS assume that the interior is constructed from the near-horizon \dof,  and that the far-away Hawking radiation is not involved.  Just to give the assumption a name, I will call it Postulate 5: the $ proximity $ postulate.

 The proximity postulate directly contradicts the statement  that  ``after a great deal of evaporation, the redundancy is between the interior and the Hawking radiation."
AMPS were of course aware of this, and an important part of their argument is devoted to  justifying  the proximity postulate.

\sc
\section{Aspects of Entanglement}

In order to make the written version of this lecture self-contained I have included a section on various aspects of entanglement.\\\

There are two situations in which large amounts of entanglement are known to occur. The first  has to do with the properties of the ordered ground states of quantum field theories including condensed matter systems. The second is almost the complete opposite; it involves entanglement that occurs as a result of complete randomness. The first case is fairly familiar; nearby subsystems tend to be highly entangled as a result of energy considerations. This type of entanglement leads to the area law for entanglement entropy, the reason being elementary; the number of lattice points adjacent to a given region is proportional to the surface area of the region. If one thinks of entanglement as the sharing of Bell pairs, then the Bell pairs in this first type of entanglement are well localized and the components of a pair are not distantly separated.

\subsection{Ordered Ground States}
\begin{figure}[h!]
\begin{center}
\includegraphics[scale=.5]{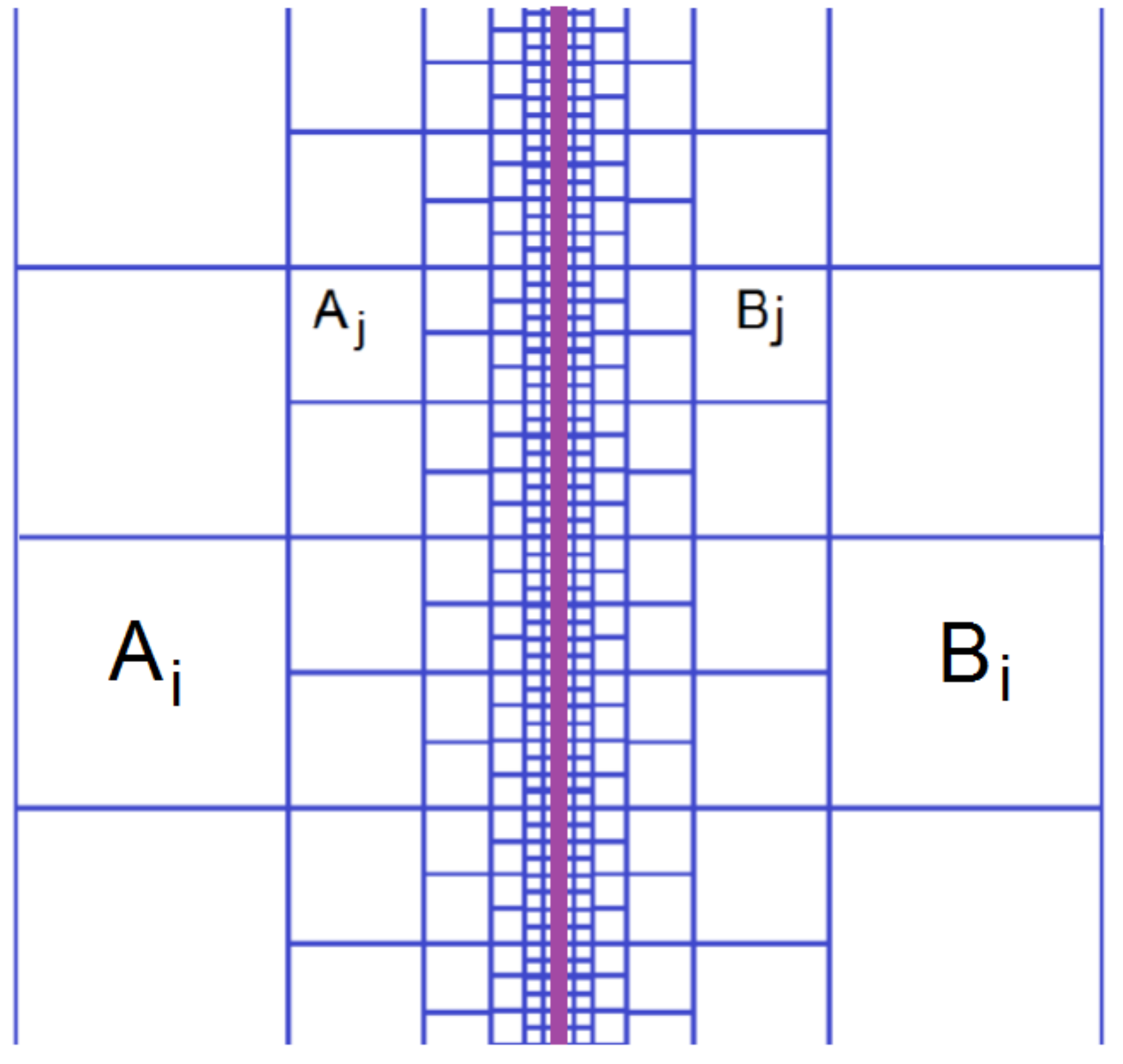}
\caption{Dividing space into two entangled half-planes. The entanglement of a conformal theory can be envisioned
in terms of mirror-image Bell pairs formed from cells on the right and left side. }
\label{grid}
\end{center}
\end{figure}

A typical example of an ordered ground state is the vacuum of a conformal field theory. If we divide space into a left and right half, the two halves will be entangled with an divergent entanglement entropy proportional to the area of the dividing plane.
\be
S= \frac{A}{l^2}
\ee
where $l$ is an ultraviolet cutoff. A rough picture of the entanglement can be provided by dividing the space on either side into cells in a scale-invariant way, as in figure \ref{grid}. The  purple line in the figure representing the  boundary between the entangled regions has been drawn thickened to represent the cutoff length $l.$

In each cell a degree of freedom can be defined by averaging the field over the cell. The degree of freedom in a cell at a distance $\rho$ from the dividing-surface  are therefore field-modes with wavelength of order $\rho.$
The entanglement across the surface can be approximated by saying that mirror image cells are entangled.
The locality and scale-invariant character of the entanglement can be roughly modeled by thinking of the cells as qubits which are entangled in Bells pairs, $A_i$ entangled with $B_i,$ as in figure \ref{grid}.  Each entangled Bell pair contributes a single bit of entanglement entropy.

\subsection{Scrambled Systems}

The other less familiar situation is entirely different in character; it occurs when energy is not a consideration at all. It is the entanglement of a scrambled system \cite{Page:1993wv}\cite{Hayden:2007cs}\cite{Sekino:2008he}. The shared Bell pairs in this type of entanglement are extremely de-localized; they are diffused over the entire system. Since it plays a large role in what follows, I will spend some time explaining scrambling entanglement. A good example is based on a random system of a large number, $N,$ of qubits. In a particular basis (sometimes called the computational basis) each qubit has two basis states labeled $0$ or $1.$

Begin with a highly non-typical state such as
\be
|\Psi_0\ra = |0000000...00\ra.
\ee
To scramble the state, randomly pick a  unitary operator $U$ from some ensemble of $2^N\times2^N$ unitary matrices. A simple ensemble is the maximally random Haar ensemble which indeed scrambles, but it is also overkill---a point we will come back to.

\begin{figure}[h!]
\begin{center}
\includegraphics[scale=.5]{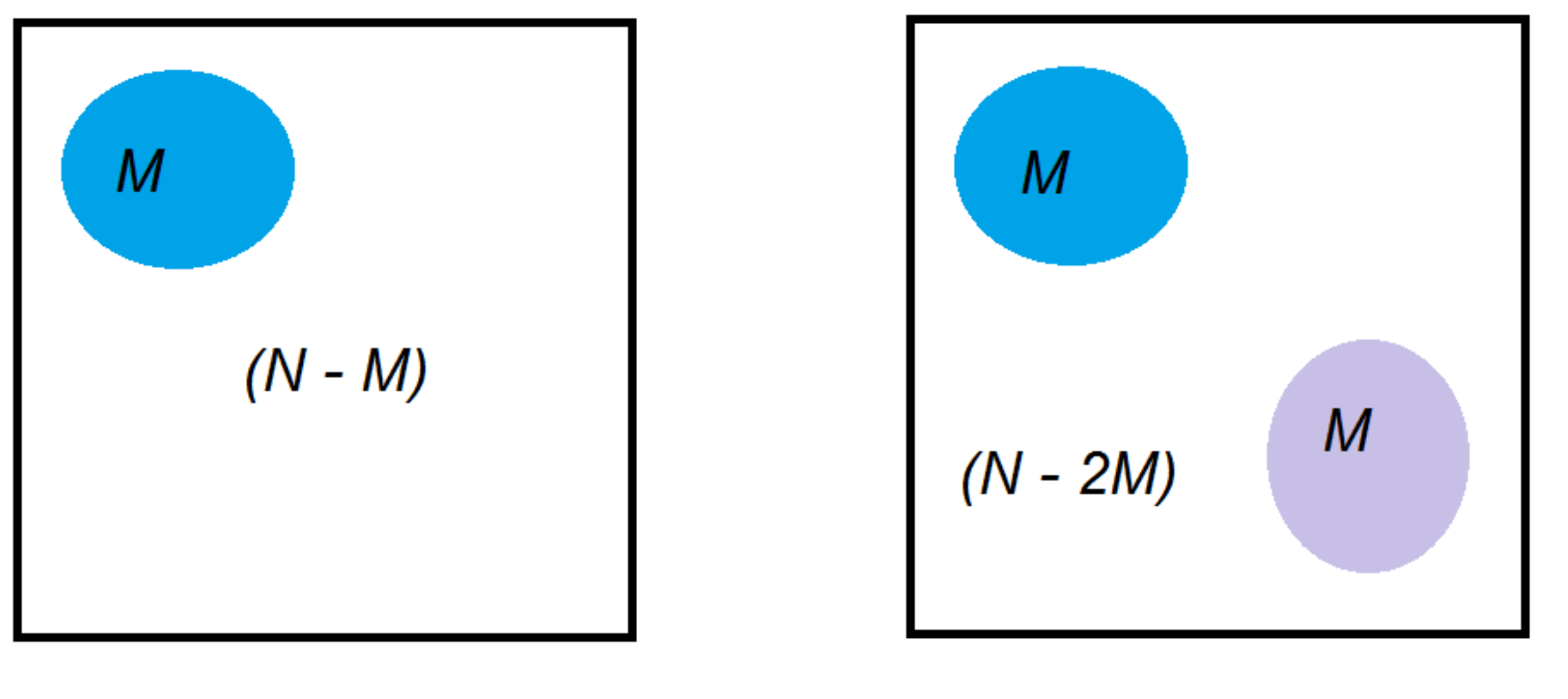}
\caption{On the left side an $N$-qubit system is divided into a small subsystem with $M$ qubits and a big subsystem with $N-M$ qubits. On the right side the big subsystem is further divided into a second subsystem with $N$ qubits and a remainder of $N-2M$ qubits. }
\label{NM}
\end{center}
\end{figure}

The scrambled  state is defined by,
\be
|\Psi \ra = U |\Psi_0\ra
\label{psi}
\ee
With overwhelming probability $|\Psi \ra$ has the  scrambled   property; namely, any small subsystem has essentially no information. A small subsystem means any subset of qubits  fewer than half the total number. If $M \ < \ N/2,$ then a system of $M$ qubits is small. In the left side of figure \ref{NM} an $N$-qubit system is divided into an $M$-qubit subsystem, and an $(N-M)$-qubit subsystem.

 The remarkable property of scrambled systems is that with overwhelming probability the amount of information in a small subsystem is negligible. The precise meaning of this statement is that for almost all matrices $U$ the entanglement entropy a small subsystem $M$ is very close to maximal,
\be
S_M = M \log 2
\label{maxS}.
\ee
(From now on I will drop the factor $\log2$ and measure entropy in bits.)
The equality sign in \ref{maxS} is not quite exact but the error is less than a single bit, and generally much less than that. I will ignore this small discrepancy in what follows.

Another way to say the same thing is that the density matrix of the small subsystem $M$ is extremely close to the maximally mixed density matrix,
\be
\rho_M = 2^{-M} I
\label{rhoM}
\ee
where $I$ is the unit matrix in the state-space of the $M$ qubit system. Again, the equality sign is correct up to negligible errors in the large $N$ limit.

It follows that the scrambled state $|\Psi \ra$ in \ref{psi} can be written in the form,
\be
|\Psi \ra = \sum_i |i\ra_s |\phi_i\ra_b
\label{entvec}
\ee
where the states $\sum_i |i\ra_s$ represent a basis for the small  $M$-qubit system, and the $|\phi_i\ra_b$ represent states in big subsystem of $(N-M)$ qubits. Moreover, the fact that the density matrix of the small subsystem is maximally mixed implies that the $|\phi_i\ra_b$ are orthonormal.

The $|\phi_i\ra_b$ are not a complete basis for the big $(N-M)$-qubit system. They only  span a subspace of dimension $2^M.$ We can think of the $|\phi_i\ra_b$ as the basis states for a subsystem of $M$ qubits that lives in the larger $(M-N)$ qubit subsystem. This subsystem is most certainly not a collection of the original defining qubits. However, it is unitarily equivalent to such a subsystem. To make this precise we can take any $M$-qubit subsystem from the $(N-M)$ system. This is shown in the right side of figure \ref{NM}. Thus we have three subsystems. The first is the original small $M$-qubit subsystem. Next is a second small subsystem which belongs to the $(M-N)$ qubit system. Finally there are the left over $N-2M$ qubits, also belonging to the big subsystem.
The point is that any vector of the form \ref{entvec} is close to a vector that can be expressed by a two step process. First define a state in which the two small subsystems are maximally entangled, and the third subsystem factors off.
\be
|\Phi \ra =\sum_i |i \ra_s |i\ra_{s^{\prime}} |00000...\ra
\ee
where
$s^{\prime}$
refers to the second small subsystem, and  $|00000...\ra$     denotes the state of the remaining $(N-2M)$ qubits. In such a state the small subsystem is manifestly maximally entangled with a subsystem of the big subsystem.

To obtain $|\Psi \ra$ from $|\Phi\ra $ we apply a unitary scrambling operator $V$ on the big $(N-M)$ qubit subsystem.
\be
|\Psi \ra = V |\Phi\ra.
\ee

The operator $V$ is the product of a scrambling operator on the big subsystem and the unit operator in the small subsystem. What $V$ does is to scramble the $M$ qubits, that are entangled with the small subsystem, and hide them among the larger $(N-M)$ qubits of the big subsystem. One point to bear in mind is that the matrix $V$  depends on the state $|\Psi \ra.$ In other words $V$ is a function of $U.$

States of this type are not special. Almost all states of the original $N$-qubit system will have this behavior, however they are divided into a small and big subsystems.

I suggested earlier that the matrix $U$ may be drawn from an ensemble which is less random than the uniform Haar measure.
A Haar-random matrix will certainly produce scrambled states, but it is very difficult to achieve Haar-randomness; indeed it can only be done by an exponential number of operations (exponential in $N$). However, scrambling can typically be achieved by a polynomial number of steps; a weaker type of randomness is sufficient. The matrix $U$ need only be drawn from the ensemble of unitary 2-designs, U2. U2-randomness is equivalent to Haar randomness for any quadratic function of the density matrix of the system, and is enough to scramble. The main point about U2 is that it is much easier to average over U2 than to average over the Haar ensemble. Averaging over the Haar-measure typically takes exponential time, while U2 averaging can be accomplished in polynomial time.

On the face of it, the two situations in which large amounts of entanglement occur; namely highly ordered ground states, and highly random scrambled states, seem to have little to do with each other. However, in black hole physics the two come together in a surprising way. The highly ordered ground state seen by a freely falling observer at the horizon is ``dual" to a highly scrambled state seen by an observer who stays outside the black hole. Moreover, the entanglements of the ordered infalling state are dual to the entanglements of the random thermal state of the exterior description.

In some ways scrambled states resemble maximally mixed states (density matrices proportional to the unit matrix), but there are subtle fine-grained differences.
For each scrambled state there are observables which distinguish it both from other scrambled states, and from mixed states. I will refer to them as fine-grained properties, by contrast with coarse-grained properties that do not encode such distinctions.

Consider a maximally impure state for the entire system described by the density matrix
\be
\rho = 2^{-N}I
\ee
where $I$ is the $2^N$-dimensional unit matrix.
The maximally impure state $\rho$ shares the property with $|\psi\ra$ that small subsystems have no information, but in the impure case,  large subsystems also have no information. Moreover $\rho$ does not  give rise to massive entanglement. No two subsystems are entangled in $\rho.$

\subsection{Coarse-grained and Fine-Grained}

In the theory of large chaotic systems there are quantities that are so ridiculously hard to keep track of that we are inclined to think of them as completely meaningless. An example would be the memory of the initial  position $x(0)$ of a particular particle in a sealed box of gas, after an exponentially long time $t$. The principles of classical mechanics imply that $x(0)$ can be expressed in terms of the coordinates and momenta of all the particles at time $t,$ but this information is of no practical value. It is so fine-grained that the tiniest perturbation in either the initial conditions or the Hamiltonian of the box will radically alter the connection between the \dof \ at time $0$ and time $t.$  We call such quantities fine-grained.
Coarse-grained quantities are the opposite; they are relatively insensitive to such tiny perturbations. For example the number of particles in a small sub-volume averaged over a one-second interval is coarse-grained.
It will be important to realize that the quantities that the AMPS argument deals with are analogous to the very fine-grained details of a box of gas.

Properties of a scrambled qubit system fall into two classes, coarse-grained properties  are sensitive to the difference between  $|\Psi \ra$ and $\rho;$ and fine-grained properties which are not. All quantities that are made out of fewer than half the qubits are coarse grained. Fine grained observables are always made out of more than half the qubits. Any test of the entanglement between small and large subsystems is fine-grained.

For the simple qubit systems we've discussed, typical fine-grained quantities are trivial. Each qubit can be described by the usual Pauli operators $\vec{{\sigma(n)}}$ where $n$ label the qubit. Any operator in the entire system can be expressed as a sum of products these Pauli operators. Suppose we consider an operator made out the  $M$ qubits of a small subsystem. Since the density matrix of any small subsystem is proportional to the identity, the expectation value of any product of the $M$ qubit operators is zero. This is  true for both a scrambled pure state and the maximally mixed state.

On the other hand when we come to operators involving more than half the qubits, things change. Consider a particular qubit--say qubit $1$---in the small subsystem of $M$ qubits that we discussed earlier. In the scrambled state that qubit is entangled with a hidden qubit in the big subsystem. Suppose we call the Pauli operators for the hidden qubit $\vec{{\tau(1)}}.$ By an appropriate choice of conventions we can assume that $\vec{{\sigma(1)}}$ and $\vec{{\tau(1)}}$ are, to a high approximation, in a singlet state, Therefore
\be
\la \Psi  | \vec{ {\sigma(1)}} \cdot \vec{{\tau(1)}}|\Psi \ra = -3
\label{-3}
\ee

But in the maximally mixed state
\be
\la \Psi  | \vec{ {\sigma(1)}} \cdot \vec{{\tau(1)}}|\Psi \ra =0.
\ee

By measuring $ \vec{ {\sigma(1)}} \cdot \vec{{\tau(1)}}$ and obtaining a number far from $-3$ one could distinguish\footnote{One can distinguish the maximally mixed state from a particular scrambled state in this manner. But one could not distinguish whether the state was pure without specifying the particular state. In other words this is not a one-shot method to determine purity.} $|\Psi \ra$ from $\rho.$

The operator $\vec{{\tau(1)}}$ is not one of the original qubit operators. It is an extremely complicated combination of all  $(N-M)$ qubits of the big subsystem, and therefore the operator in \ref{-3} is made out of more than half the original qubits. From the fact that the expectation value is different in the scrambled and mixed states we recognize it as representing a fine-grained property.

Fine-grained operators are extremely dependent on the scrambled state in the following sense. If we pick the particular combination of original qubit operators that define $\vec{{\sigma(1)}} \cdot  \vec{{\tau(1)}}$ for the state $|\Psi\ra$ and evaluate it in another scrambled state $|\Psi^{\prime}\ra$ the result will be negligible (exponentially small) \cite{patrick}. So to test out if a state is pure by measuring a fine-grained operator, one has to know in advance exactly what scrambling dynamics has taken place.

Fine grained quantities for large systems generally don't play any role in practical many-body physics. The quantities that interest us usually don't depend on whether a large system is in a pure state or a thermal ensemble unless the system is in the ground state. Fine-grained observables are much too difficult to measure. Thus we have very little experience with fine-grained physics, but the questions that will occupy us in this lecture are of the most fine-grained kind.

\subsection{Distillable Entanglement}
\subsubsection{Pure States}

It's important to have a quantitative concept of how entangled two subsystems are. We imagine a system of qubits and consider two subsystems will be denoted by \b \ and \h.
To define the amount of entanglement between \b \ and \h \  I will introduce a concept from quantum information theory  called \it distillable entanglement \rm \cite{Plenio:2007zz} represented by the symbol $D.$  I will not attempt to be too precise in its definition;  in essence it is the number of Bell pairs shared by two subsystems.

First of all, if we insist on exactly maximal entanglement, then the generic answer is zero, but if we relax the tolerance a bit, the generic answer is lots. An exact Bell pair is a pair of qubits, $B$ and $H_{B},$ such that:

\bigskip

1) The density matrix of the union of the two is pure. In other words the Von Neumann entropy of the union $B\cup H_B$ is exactly zero.

2) The density matrices of the individual subsystems $B$ and $H$  are each maximally random, i.e., proportional to the identity matrix. We can also say that the Von Neumann entropy of each subsystem is maximal and equal to $1.$

\bigskip

To relax these conditions we can introduce a small parameter $\epsilon$ and require that:

\bigskip

$1^{\prime}$) The density matrix of the union of the two is almost pure. In other words the Von Neumann entropy of the union $B\cup H_B$ is less than $\epsilon$.

$2^{\prime}$)  The density matrices of the individual subsystems are almost maximally random. The Von Neumann entropy of each subsystem is almost maximal and greater than $1-\epsilon.$

\bigskip

This defines a regulated version of a Bell pair.

\subsubsection{The Page Transition}
Consider a scrambled system of $N$ qubits, and divide it into two subsystems with $M$ and $(N-M)$ qubits.
The following is true : Given any small $\epsilon,$ there is an $N(\epsilon),$ such that if the total number of qubits is greater than $N(\epsilon),$ then the number of regulated Bell pairs is equal to the smaller of $M$ and $(N-M)$.
A more general way to express this is that the number of approximate Bell pairs is equal to the entanglement entropy of of the subsystems. For a large black hole the numbers are so big that the difference between regulated and exact Bell pairs is not important.

\begin{figure}[h!]
\begin{center}
\includegraphics[scale=.3]{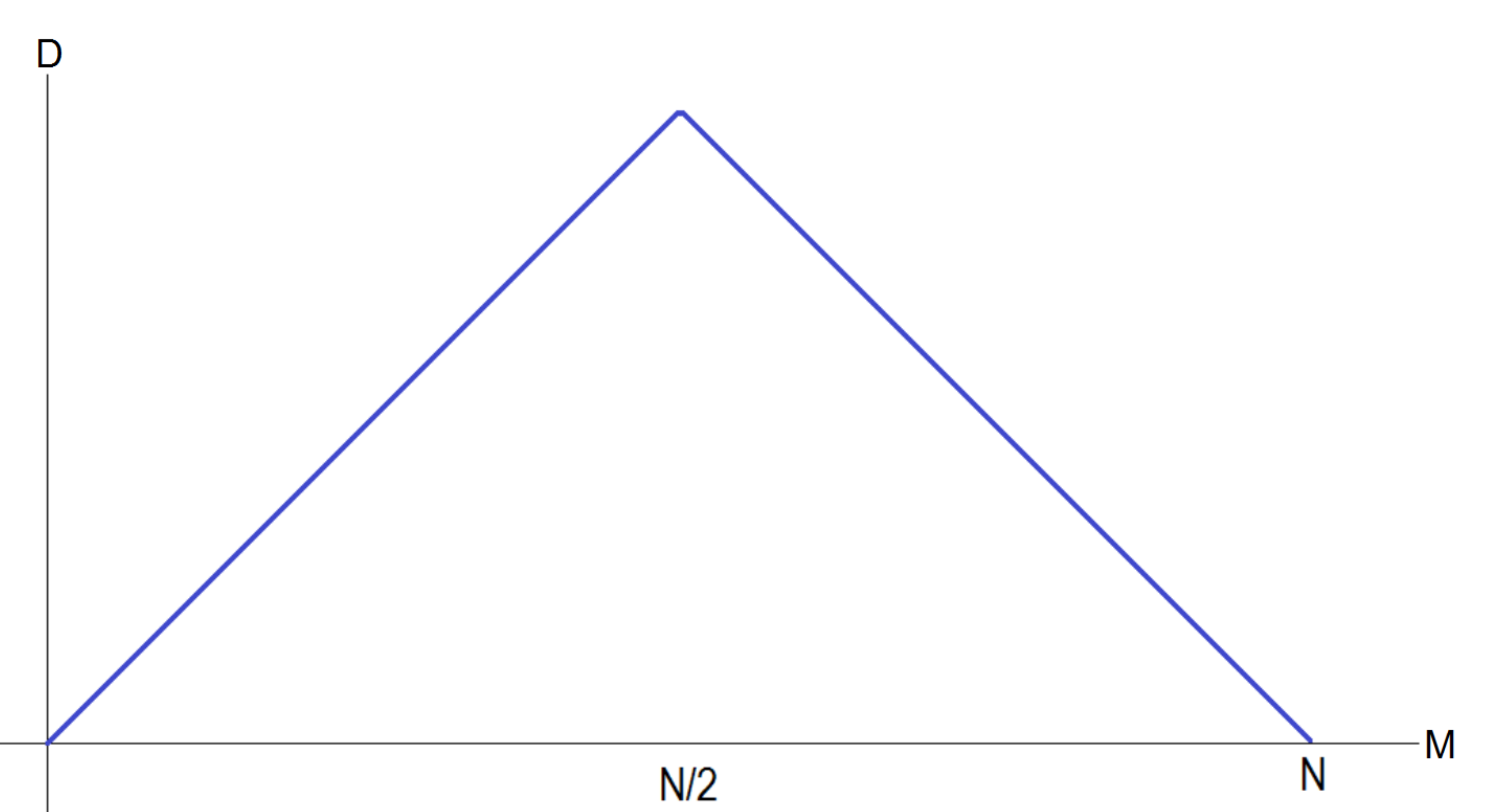}
\caption{ The Page curve for distillable entanglement.       }
\label{g3}
\end{center}
\end{figure}

In figure \ref{g3} the distillable entanglement is plotted as a function of $M.$
As $M$ increases from zero, the number of Bell pairs increases linearly until it reaches a maximum at $M=N/2.$
At that point the curve  exhibits a sudden change in the slope, and by symmetry, it decreases to zero when $M=N.$ This sharp transition or cusp-like behavior was first observed by Page \cite{Page:1993wv} and the curve is sometimes referred to as the Page curve.

\subsubsection{Mixed States}
Let's consider the distillable entanglement between \h \ and \b, assuming the total system is not in a pure state.  Counting the distillable  entanglement is more difficult if the bipartite system under consideration is not pure, because the concept of entanglement entropy does not exist. Nevertheless the distillable entanglement $D$ is defined, for our purposes  as follows:

Consider a unitary operator $U$ constructed as the tensor product of a $2^{N_H}\times 2^{N_H}$ matrix in the Hilbert space of the subsystem \h, and the identity matrix in \b.  Apply it to the density matrix $\rho_{BH}$ of the \h\b \ system. The result is $\hat{\rho}_{BH}.$
\be
\hat{\rho}_{BH} = U^{\dag}\rho_{BH} U
\ee
Next, pair the \b-qubits  with a subset of the \h-qubits  and count the number of regulated Bell pairs. Finally maximize that number with respect to all $2^{N_H}\times 2^{N_H}$ unitary transformations. Basically what we are doing is unscrambling the \h \ system and counting the Bell pairs. The process is called distilling and the resulting number of Bell pairs is the distillable entanglement\footnote{There are several notions of distillable entanglement including various ``one-shot" definitions. Generally they all allow transformations on both subsystems. Since the definition I am using restricts the search for Bell Pairs by allowing only transformations on \h, it gives a smaller answer than other definitions. }.

The distillable entanglement is very difficult to compute but we can bound it, and in some situations of interest the bound is almost saturated. The useful bound on $D$ is given in terms of the mutual information of the \h\b \  system. If $S_B, \ S_H,$ and $S_{BH}$ are the Von Neumann entropies of \b, \h, and \h\b \ (the union of \h \ and \b) we define
\be
\mu = \frac{1}{2}(S_B + S_H - S_{BH}   )
\ee
(Note that $\mu$ is defined to be half the usual mutual information.)

Suppose that  the \h\b \ subsystem is purified by a third subsystem \r.  Then it follows that $S_{BH} = S_R$ and we can write
\be
\mu = \frac{1}{2}(S_B + S_H - S_{R}   )
\label{mu=sb+sh=sr}
\ee

It is known that distillable entanglement is bounded by $\mu$ \cite{Plenio:2007zz},
\be
D \leq \mu.
\ee

Another simple fact that helps in computing $\mu$ is that for a scrambled system the Von Neumann entropy of a small subsystem is always maximal \cite{Page:1993wv}. This means that for any subsystem of $M$ qubits, its entropy is given by $M $ as long as $M<N/2.$

Finally, there are two situations in which $D$ equals $\mu$ or is very close to it. The first case is $\mu = 0.$ Since both $\mu$ and $D$ are never negative it follows that if $\mu$ vanishes, so does $D.$

The second less trivial case is when $\mu$ is maximal or close to it. This happens when $\mu \approx M.$ In that case $D\approx M .$

These facts will be helpful in explaining the firewall argument.

\sc
\section{Complementarity and The Firewall Argument}

\subsection{Degrees of Freedom}

\begin{figure}[h!]
\begin{center}
\includegraphics[scale=.5]{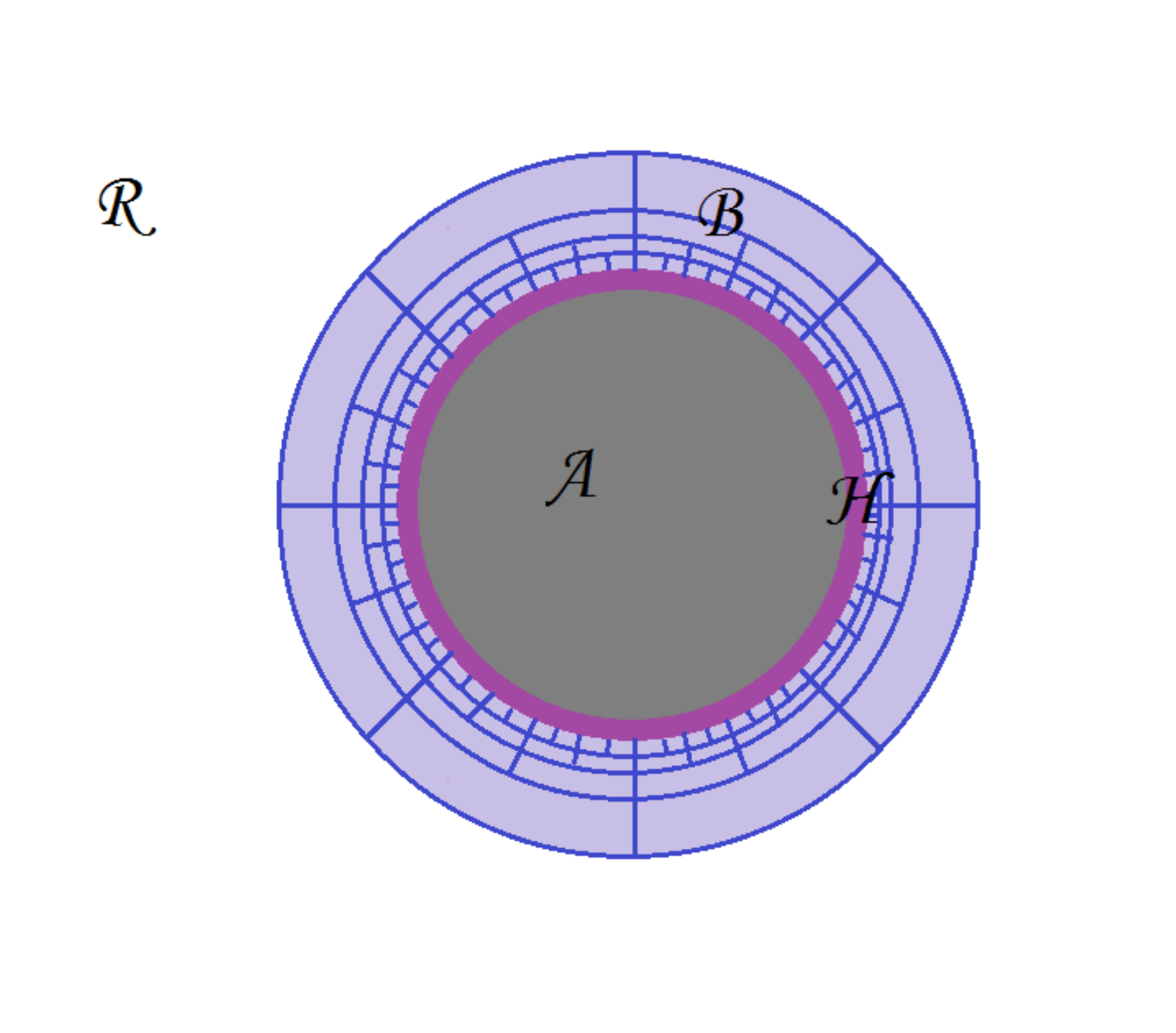}
\caption{The Black hole geometry is divided into four regions, \r, \b, \h, and \a. Region \b \ is shown divided into thermal cells, each with a single bit of entropy. }
\label{0}
\end{center}
\end{figure}

Let's begin with an account of the firewall argument as described in \cite{Susskind:2012uw}. The argument, which assumes the proximity postulate, is
 based on a simplified model of a black hole that starts by dividing the black hole geometry into four  regions as in figure \ref{0}. The first three called \r, \b, and \h \ form the exterior of the black hole; in other words the regions outside the horizon.  The most distant region lies beyond the photon sphere at $r=\frac{3R_s}{2}$  ($R_s$ is the \S \ radius). This outer region is called \r \ for radiation. Detectors in \r \ unambiguously detect Hawking radiation. The degrees of freedom in \r \ are very low energy and low angular momentum massless quanta.

Moving inward, the \it zone \rm labeled \b \ is the next region,  which lies between the photon sphere and the stretched horizon. This region is well-approximated by Rindler Space. The \dof \ of \b \ are quantum fields but with a short-distance cutoff at the string scale $l_s,$ where string effects become important. The zone lies between the photon sphere and a proper distance $l_s$ from the horizon.

The remaining portion of the black hole exterior is the stretched horizon  labeled \h. The \dof \ of \h \ are not field theoretic; the main thing I will assume  is that they are fast scramblers\cite{Page:1993wv} \cite{Hayden:2007cs} \cite{Sekino:2008he}. Possibly they form a matrix system  as in Matrix theory \cite{Banks:1996vh}.

In the static \S \ frame, unlike the infalling frame, the zone \b \ and the stretched horizon \h \ are in thermal equilibrium at a non-zero temperature. The dimensionless  coordinate temperature is $\frac{1}{2 \pi} $ and the local  proper temperature, in the zone, varies according to
$$T=\frac{1}{2\pi \rho}.$$
Here $\rho$ is the proper distance from the horizon. At the stretched horizon the proper temperature is $\frac{1}{2\pi l_s}.$

The fact that the \h\b \ system is thermalized is the basis for the random qubit model. Pure states of a complex system which have a finite energy per degree of freedom are a lot like scrambled states. To make the connection explicit, one can subdivide a thermal system into cells, in such a way that each cell has about one bit of entropy. For low temperature systems the thermal cells are large, while for hot systems they are small. If the system is rescaled so that there is one thermal cell per unit volume, then the temperature will be rescaled to order unity, and each thermal cell can be thought of as a single qubit of a scrambled system.

As we will see, the similarity between one side of figure \ref{grid} and the cells of figure \ref{0} represent the duality between the ground-state entanglements of a pure state and the scrambled entanglements of the thermal state describing the \h\b \  system .

The total entropy of the black hole includes the entropy in \h \ as well as that in \b. The entropy in field-modes of \b \ can be computed. It would be divergent without the cutoff, but with a cutoff at $\rho = l_s$ it's of order
\be
S_B \sim  \frac{R_s^2}{l_s^2}.
\ee
 We can also write it as \cite{Susskind:1993ws},
\be
S_B = g^2 S_{BH}
\ee
where $g$ is of order the four-dimensional string coupling constant, and $S_{BH}$ is the total black hole (Bekenstein Hawking) entropy. If we assume $g<<1$ then zone entropy is a small numerical  fraction of $S_{BH}.$

If the black hole is formed by collapse, then  while it is young the radiation \dof \ can be ignored. But as it evaporates entropy is transferred from the \h \b \ system to \r. For simplicity we can assume that the total entropy is conserved although in practice it increases by a factor of about $3/2$ \cite{Page:1983ug}.

In visualizing the field theory modes in the zone, it's helpful to transform from proper distance $\rho$ to the tortoise coordinate $u$ where $$u= \log{\frac{\rho}{l_s} } .$$  The horizon is at $u=-\infty$ but the stretched horizon is at $u=0.$ Thus in the zone  the physical range of $u$ is from zero to the photon sphere at $u=\log{\frac{2MG}{l_s}}.$
For simplicity  consider a massless scalar field $\phi$. If the field is weakly interacting then we can express it in terms  individual angular momentum modes in which case the wave equation in the zone takes the form
\be
\nabla^2 \phi + \frac{l_s^2}{R_s^2}
L^2 e^{2u} \phi =0
\ee
 where: $R_S$ is the \S \ radius of the black hole: $l_s$ is the string length scale: $L $ is the angular momentum of the mode. The expression $\frac{l_s^2}{R_s^2} L^2 e^{2u}$
 is the usual centrifugal barrier which inside the photon sphere is attractive.

 In tortoise coordinates the dimensionless coordinate temperature is $\frac{1}{2\pi}$ and the thermal wavelength $\Delta u$ is order $1.$ The modes that are  excited in the equilibrium state of the black hole can be classified by angular momentum, and by the radial tortoise coordinate $u.$ The $u$-axis can be coarse-grained into thermal cells with a spread of order  $\Delta u =1.$

 For a given angular momentum the range of $u$ runs from $u=0$ at the stretched horizon, to a value determined by the centrifugal barrier. Using the fact that the temperature is of order unity, the modes become frozen out of the thermal ensemble when $e^u> \frac{R_s}{L l_s}.$ Thus each $L$ mode lives in a ``box" defined by
 \be
 0<u<   \log{\frac{R_s}{l_s L} }
 \ee
It follows that for each total angular momentum $L$ there are $(2L+1) \log{\frac{R_s}{l_s L} } $ effective field modes in the zone, and each field mode carries about a single bit of entropy. Roughly speaking each of these modes is a qubit. The qubits can be thought of as each inhabiting a thermal cell as in figure \ref{0}.

The simplified model represents the black hole as a system of $N$ qubits where $N$ is the Bekenstein-Hawking entropy shortly after collapse.  The qubits are assigned to the subsystems \h, \b, and \r \ according to,
\be
N = N_H + N_B + N_R.
\ee
At any given time the black hole entropy is
\be
S_{BH} = N_H + N_B
\ee
and the entropy in the radiation is\footnote{This formula is correct for the coarse-grained entropy of the radiation. For the fine-grained Von Neumann entropy, it is correct up to the Page time. }

\be
S_R = N_R.
\ee
The fraction of the black hole degrees of freedom carried by \b \ is
\be
\frac{N_B}{S_{BH}} = g^2.
\ee
We can assume that the black hole is formed in a pure state and that the internal dynamics  quickly scrambles it, long before any appreciable amount of radiation has been emitted.

These qubit modes do not describe everything that can happen in the zone. For example, a particle can have an
energy much higher than the thermal energy $\sim \frac{1}{\rho}.$ Such modes describe infalling particles that fall in from infinity with energy larger than the Hawking temperature, but they contribute a negligible amount to the entropy of the black hole. When a particle with high energy falls in, in the static frame it disturbs the equilibrium for a short time. But in the scrambling time of order $R_s \log{R_s}$ the energy is distributed into the  thermalized modes of the stretched horizon \h.

The model that I'll use for evaporation is  very simple. One by one, qubits are transferred from the black hole subsystem (\h \b) to the radiation subsystem. The entire system remains in a pure state but the \h \b \ subsystem loses its purity. Eventually all the qubits are transferred to \r \ and the black hole disappears. The final state of the radiation is pure, but it is highly scrambled.

The fourth region of the black hole is the interior---the region behind the horizon---called \a. As long as we keep away from the singularity, the degrees of freedom in \a \ are similar to those in \b; they are field theoretic with a  cut-off at $l_s.$ However, the main point of
BHC is that they are not new independent \dof. The \dof \ in \a \ are constructions built out of the exterior \dof \ in \h, \b, and \r.

\subsection{Entanglement of \a \ and \b }

\begin{figure}[h!]
\begin{center}
\includegraphics[scale=.3]{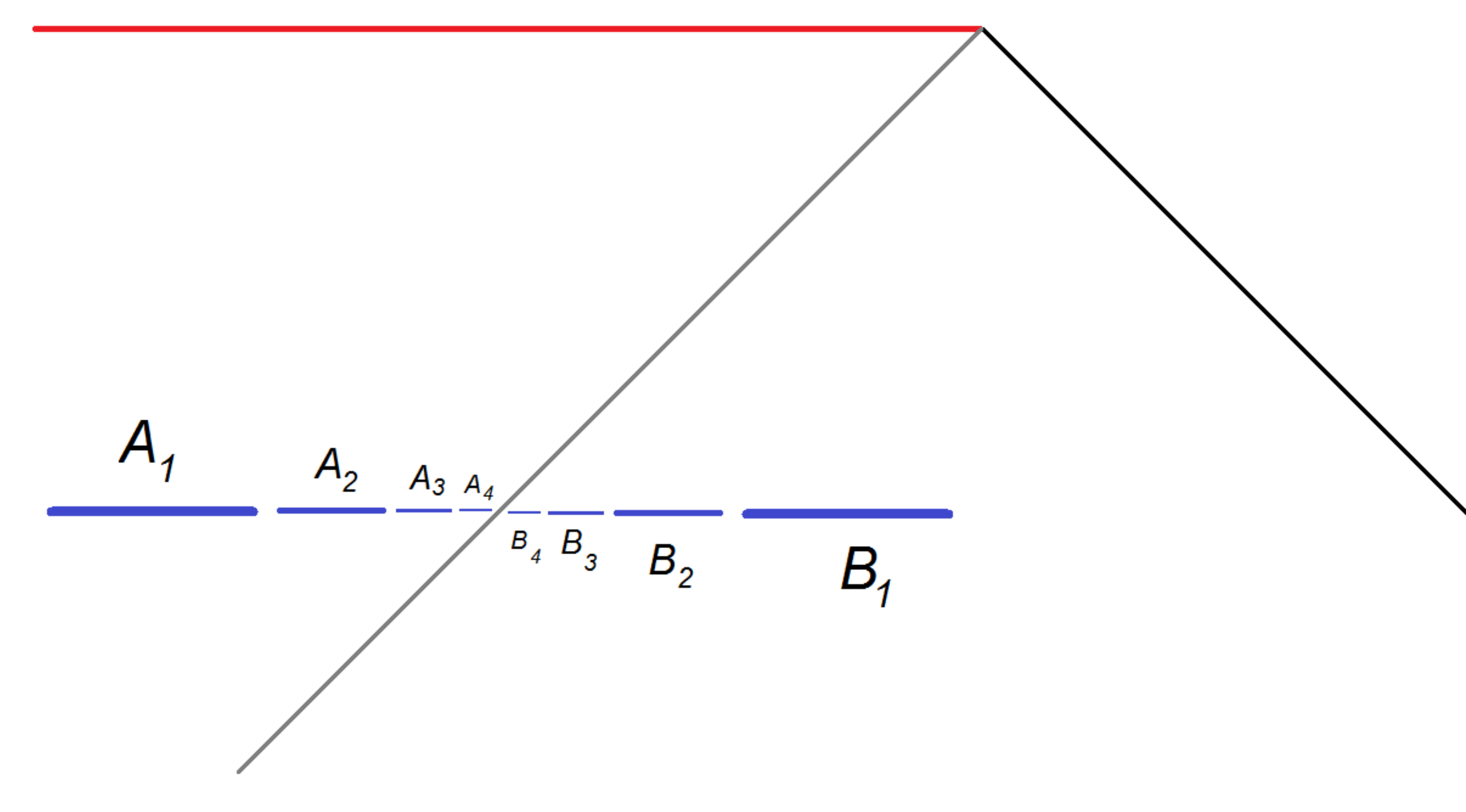}
\caption{The pairing of $A$ and $B$ modes can be carried out on a space-like surface in an infalling frame.          }
\label{1}
\end{center}
\end{figure}

To make their argument AMPS take the perspective of a freely infalling observer passing from the zone to the interior. According to BHC such an observer sees ``no drama" at the horizon, meaning that she sees the ordered Minkowski vacuum of a field theory like the one in Section 2.1. The field modes in \a \ and \b \ must be entangled in approximate Bell pairs---an example of the ground-state entanglement dictated by energy considerations. Breaking the entanglements would lead to a large energy density in the infalling frame,

As we have seen, a mode in \b \ can be characterized by a  spherical harmonic and a radial tortoise coordinate, uncertain to about $\Delta u = 1.$ The same is true of the modes in \a. In fact the \b \ modes and \a \ modes come in matched pairs of opposite angular momentum and similar tortoise distance from the horizon. This pairing can is displayed in  figure \ref{1}.

Let's denote a particular mode in \b \ by the notation $B_i$ and the corresponding partner in \a \ by $A_i.$ The entanglement that AMPS assumes is between $A_i$ and $B_i.$ For simplicity I will follow AMPS's idealized assumption that the $A$ and $B$ can be treated as qubits and that in the infalling frame $A_i$ is maximally entangled with its partner $B_i.$ In other words, in the infalling frame the \a \b \ system consists of a collection of $g^2 N$ maximally entangled Bell pairs. AMPS explain that any significant disturbance to the entanglement of these Bell pairs constitutes a violent perturbation of the infalling vacuum, and would certainly destroy an infalling observer. Therefore the no-drama postulate \cite{Almheiri:2012rt} of BHC requires that the $A,B$ pairs remain in a state of maximal entanglement as the black hole evaporates\footnote{Entanglement is a necessary requirement but not sufficient. There are many entangled states but only one of them has the correct form to describe a smooth space between $A$ and $B$.}.

This discussion of $A,B$ entanglement in the infalling frame must be translatable to the language of the exterior \dof, since by assumption, the interior \dof \ are constructed from the exterior \dof. The exterior description is thermal, and by appropriately defining thermal cells, it can be thought of as a scrambled system. That is why I said earlier that the ordered entanglement of a ground state is dual to the scrambled entanglement of a random (thermal) system.  Duality between infalling ground state and exterior thermal state is of course not new, but the point I want to emphasize is the duality of two kinds of entanglement: the entanglement of ordered ground states, and the entanglement implicit in scrambled states.

Since most of the exterior \dof \ are in \h, we can assume that the $A_i$ are constructs made of the \h \ qubits. Identifying  $A_i$ in \h \ is a matter of finding a unique subsystem of \h \ which is maximally entangled with $B_i$ (the partner of $A_i$). In general there is no guarantee that such a subsystem of \h \ exists. However, for the case of a relatively young black hole we can be sure that it does.

 By relatively young I mean that the black hole has already scrambled, but the evaporation is negligible. In that case the \h\b \ system is in a pure but scrambled state. In other words;
to a high degree of approximation, every small subsystem is described by a random density matrix proportional to the identity, but the overall state is pure.

 Given that the state of \h\b \  system is pure, there is an important consequence of this fact; namely, that to an equally
high degree of approximation, every small subsystem is maximally entangled with the rest of the system.
 Since \b \ is a small subsystem of the \h\b \  system, it follows that \b \ is maximally entangled with \h. Furthermore each qubit of \b \ is almost exactly maximally entangled with a unique subsystem of \h.

 \h \ may be given to us as some sort of recognizable quantum system, such as matrix quantum mechanics  \cite{Banks:1996vh}. But the
subsystem of \h \ that $B_i$ is entangled with is unlikely to be  easily recognizable from the defining \dof \ of \h.
Scrambled systems hide their entanglements in extremely difficult codes. Typically the code involves a  unitary \it descrambling transformation \rm acting on the \h \ subsystem. The transformation descrambles the hidden qubits that are entangled with specific qubits of \b. Nevertheless, the purity and scrambled nature of the \h\b \ state is enough to insure that each $B_i$ is partnered with a subsystem,
 $H_{B_i},$ of the stretched horizon. (The notation means that  $H_{B_i}$  is a subsystem of \h,  and that it is maximally entangled with $B_i.$) By the monogamy of entanglement  $H_{B_i}$  is unique.

 The conclusion of this line of argument is obvious.

 1) In the infalling frame $B_i$ and $A_i$ are maximally entangled.

 2) In the exterior frame $B_i$ and   $H_{B_i}$ are maximally entangled.

 3) Maximal entanglement is monogamous.

 Therefore it follows that
 $A_i$ and $H_{B_i}$ must be the same thing. The formal equation
 \be
 A = H_B
 \label{A=hb}
 \ee
 expresses this identification.

Perhaps another way to say this is that \h \ is the hologram at the horizon, that  represents the interior \a. It is evident that the relation between the interior and exterior of the black hole is extremely fine-grained from the point of view of the exterior degrees of freedom. I'll refer to it as the \h $\Longleftrightarrow$ \a \ mapping.

\subsection{Non-linearity of \h $\Longleftrightarrow$ \a \ Mapping}

An issue that is bound to come up, is the non-linearity of the \h $\Longleftrightarrow$ \a \ mapping\footnote{I thank Raphael Bousso and Douglas Stanford for discussions about this point.}. By non-linear I mean that the relation between  $A_i$ and operators in \h \  depends on the initial state $|\Psi_0\ra.$ That's because the particular form of $H_{B_i}$ is state-dependent. Although this does not imply an observable non-linear violation of quantum mechanics in either the exterior or infalling frames, it does seem to violate the linear  spirit of quantum mechanics.

Such non-linearity of the  \h $\Longleftrightarrow$ \a  \ mapping is  not completely new. It occurs in the  simple \it pull-back---push-forward \rm strategy \cite{Freivogel:2004rd}  for young black holes \cite{Susskind:2012uw}.
Suppose that the black hole is made by sending in a shell of a particular composition. The shell could be a coherent electromagnetic wave, a similar gravitational wave, or a mix of the two.

To carry out the pull-back---push-forward procedure, the operator $A_i$ has to be pulled back to the remote past using the low energy  equations of motion in the infalling frame. Since the equations of motion have to be pulled back through the shell, the operator one obtains in the remote past will depend on the nature of the shell. That dependence will also be present after the operator is pushed forward. Therefore the dictionary between interior operators and operators in the Hawking radiation depends on the initial state through the dependence on the state of the shell. This is a mild form of the same kind of non-linear dependence.

Non-linear dependence in the \h $\Longleftrightarrow$ \a \ mapping is probably inevitable. However, linearity can be restored by embedding the system in a larger system. For example, if the Hilbert space is big enough to include the operators which create the shell, then instead of saying the map depends on the state, we would say that the operator $A_i$ maps to an exterior operator in the joint Hilbert space of \h \ and the additional factor describing the shell.

\subsection{Evaporation}

Assuming the proximity postulate, a necessary condition for an uncorrupted black hole interior is that the distillable entanglement between \b\ and \h \ should be equal to the number of qubits in \a. If the number is less than that, then there is not enough of an entanglement resource to define all the interior modes. Even worse, if $D=0$ it is impossible to define any vacuum modes in \a. That's the case in which AMPS argue that the geometry is terminated at the horizon by a firewall. The AMPS argument can be formulated as a calculation which shows that the \h\b \ distillable entanglement goes to zero before the black hole has evaporated.

I will adopt the  very simple model of  evaporation as in Page's description \cite{Page:1993wv}, and the one used by Hayden and Preskill \cite{Hayden:2007cs}. Represent the collection of $N$ qubits by a box, subdivided into three boxes \h, \b, \ and \r. At first the \r-box is empty and the $N$ qubits occupy the boxes \b \ and \h.
The \b-box has $g^2 N$ qubits and the \h-box has $(1-g^2)N.$ The total box is scrambled.

\begin{figure}[h!]
\begin{center}
\includegraphics[scale=.3]{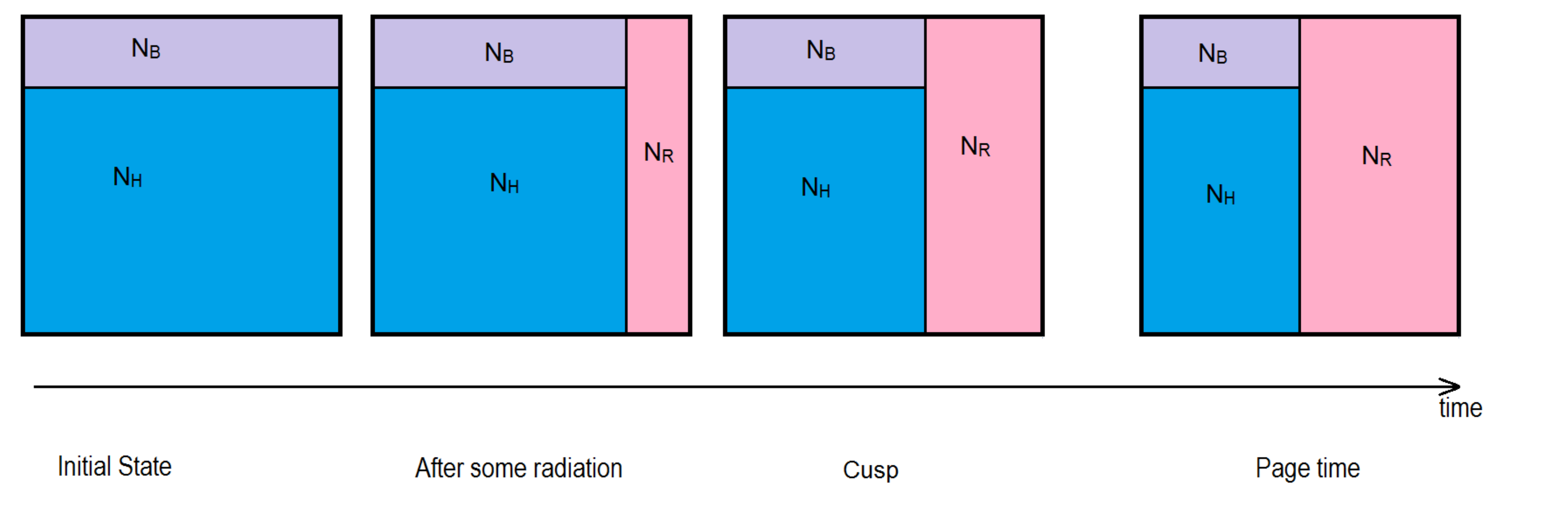}
\caption{  Initially $N$ qubits are distributed into boxes \h \ (blue) and \b (purple). As time evolves qubits get transferred to \r (pink).        }
\label{2}
\end{center}
\end{figure}

As the evaporation proceeds the \b \ and \h \ boxes shrink, while retaining the same relative size. The decreasing number of qubits is compensated by a growing number in the \r-box, the total number being kept fixed. Schematically this is shown in figure \ref{2}.

Obviously the state of the \h\b  \ system does not remain pure as qubits are transferred to \r. The question is what happens to the entanglements between \b \ and \h? Does the existence of matched Bell pairs persist so that we can continue to identify the  $A_i$ with $H_{B_i}?$

\subsection{Loss of \h, \b \ Entanglement}

In the infalling frame \a, \b \ entanglement is essential for the no-drama postulate, i.e., for the existence of a smooth horizon. Translated to the exterior description, \h, \b \ entanglement is essential. How much entanglement? In the infalling frame the number of Bell pairs is equal to $g^2 S_{BH}.$ As we will see, that amount of entanglement persists for a long time as the black hole evaporates. But at some point the entanglement begins to diminish, and by the Page time it vanishes. This is the crux of the AMPS argument.

Armed with the facts of Section 2.3, it is easy to prove the following:

\bi
\item As the black hole evaporates and $N_R $ increases, the distillable entanglement between \b \ and \h \ remains maximal and equal to $N_B,$ until a specific ``cusp" time $t_c.$ As a proportion of the black hole entropy the fractional distillable entanglement satisfies
    \be
    \frac{D}{S_{BH}} = g^2 \ \ \ \ \ \ \ \ \ (t< t_c)
    \label{D=g^2}
    \ee

    To prove this we use \ref{mu=sb+sh=sr} and observe that as long as \h \ is greater than half the system, then
    \bea
    S_B \eq N_B  \cr \cr
    S_R \eq N_R \cr \cr
    S_H \eq S_B + S_R
    \eea
    which gives
    \be
    \mu_{BH} = N_B
        \ee
    Since this is the maximal value for $\mu_{BH}$ we can also write
$$
  D_{BH} = N_B
   $$
   or
    \be
    \frac{D}{S_{BH}} = \frac{N_B}{N_B+N_H}
    \label{D=g^22}
    \ee
    which is equivalent to \ref{D=g^2}.

   \item    This behavior continues until the point where $N_H$ is half the total number of qubits. The cusp time $t_c$ is defined by the condition that $N_H$ is half the total number of qubits $N.$ One may also write that at the cusp time $N_R = N_c  \equiv N_H - N_B.$

       Note that the cusp time is earlier than the Page time $t_p$ at which $N_R$ becomes half the system.

   \item After the time $t_c$  the fractional distillable entanglement decreases linearly with time. It  vanishes at $t_p$
   and stays equal to zero until the black hole evaporates. To see this we note that between the cusp time and the Page time all three subsystems have less than half the total number of qubits. Therefore $S_B = N_B,$ \  $S_H = N_H,$ and $S_R = N_R.$ It follows that
   \be
   \mu_{BH} = \frac{1}{2}(N_B + N_H - N_R)
   \ee
   which can be written in the form
   \be
   \mu_{BH} = N_B -\frac{N_R -N_c}{2}
   \ee
In other words the mutual information begins to decrease relative to $N_B$ once the cusp is passed. It is easy to see that it vanishes when $N_R = N_H + N_B,$ i.e., at the Page time.

From the fact that $\mu$ bounds $D$ we see that the distillable entanglement between \h \ and \b \  also decreases to zero at the Page time. \\
\ei

The first fact indicates that $D$ remains large enough so that the interior degrees of freedom can be defined for a long time.  Indeed for small $g$ the value of $t_c$ is almost the Page time. This is good news for  a long-lived interior geometry, but we should be clear: distillable entanglement is a necessary condition for an uncorrupted region \a, but it may not be sufficient. It is quite possible for $A_i$ and $B_i$ to be maximally entangled but in the wrong Bell state.

But after the Page time there is no hope; the fine-grained quantities associated with \h\b \ entanglement have disappeared altogether.
 As long as we insist that the interior be built from near-horizon \dof \ the evaporation will destroy the necessary entanglements and a firewall must replace the smooth horizon, at least if the proximity postulate is correct.

\begin{figure}[h!]
\begin{center}
\includegraphics[scale=.3]{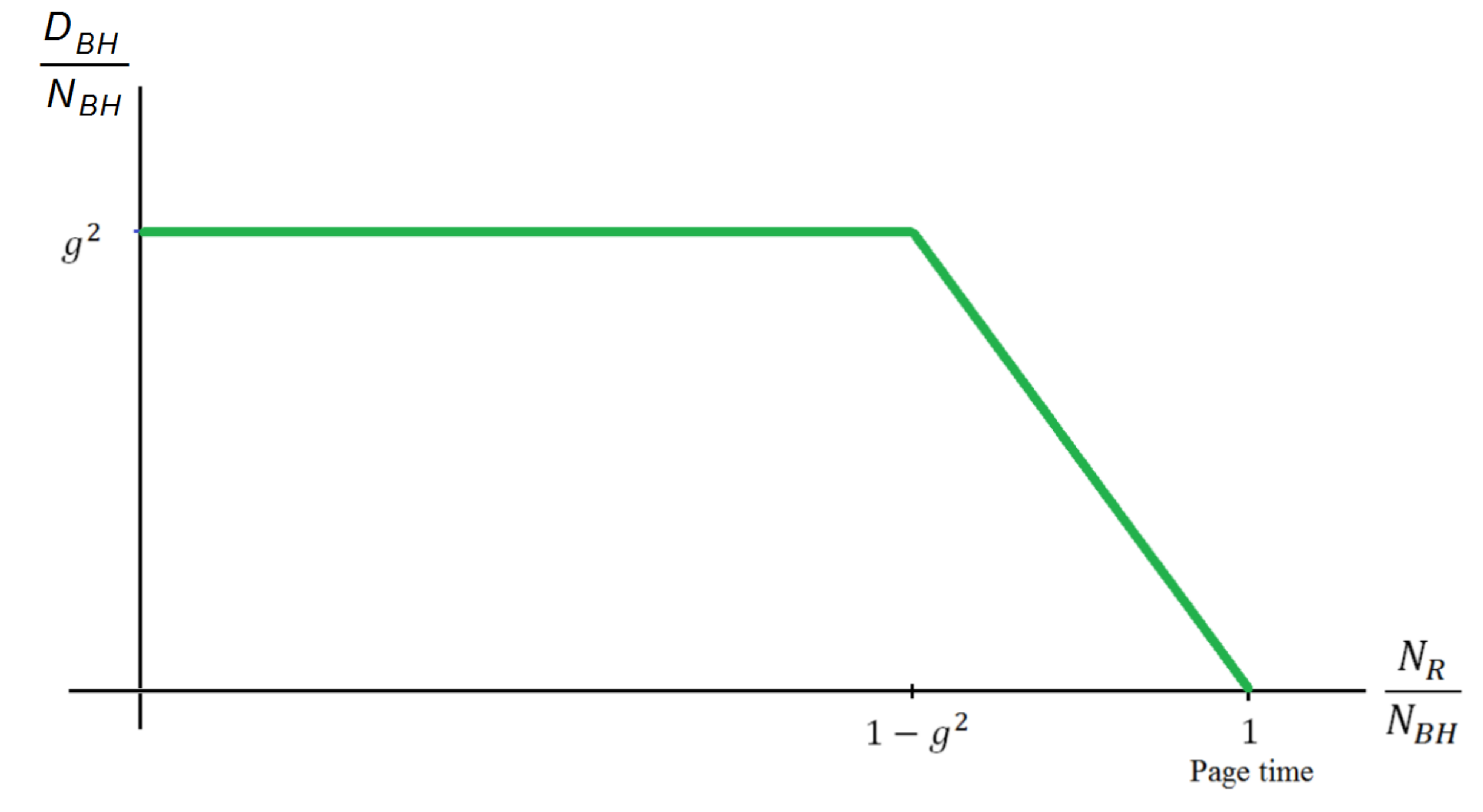}
\caption{  The \h\b \ fractional distillable entanglement stays constant until the cusp time. It then decreases
until it vanishes at the Page time.        }
\label{3}
\end{center}
\end{figure}

The evolution of the \h\b \ distillable entanglement is illustrated in figure \ref{3}.

\sc
\section{ $\cal{A} = R_B?$}
\subsection{Redundancy of \a \ and \r }

One can argue that AMPS did not prove that the standard postulates of complementarity are inconsistent, but only that they are inconsistent with the proximity postulate. Turning it around,  they proved that the first four postulates predict that the proximity postulate must be wrong, and that  information in \a \ must eventually become redundant with information in \r.

\begin{figure}[h!]
\begin{center}
\includegraphics[scale=.3]{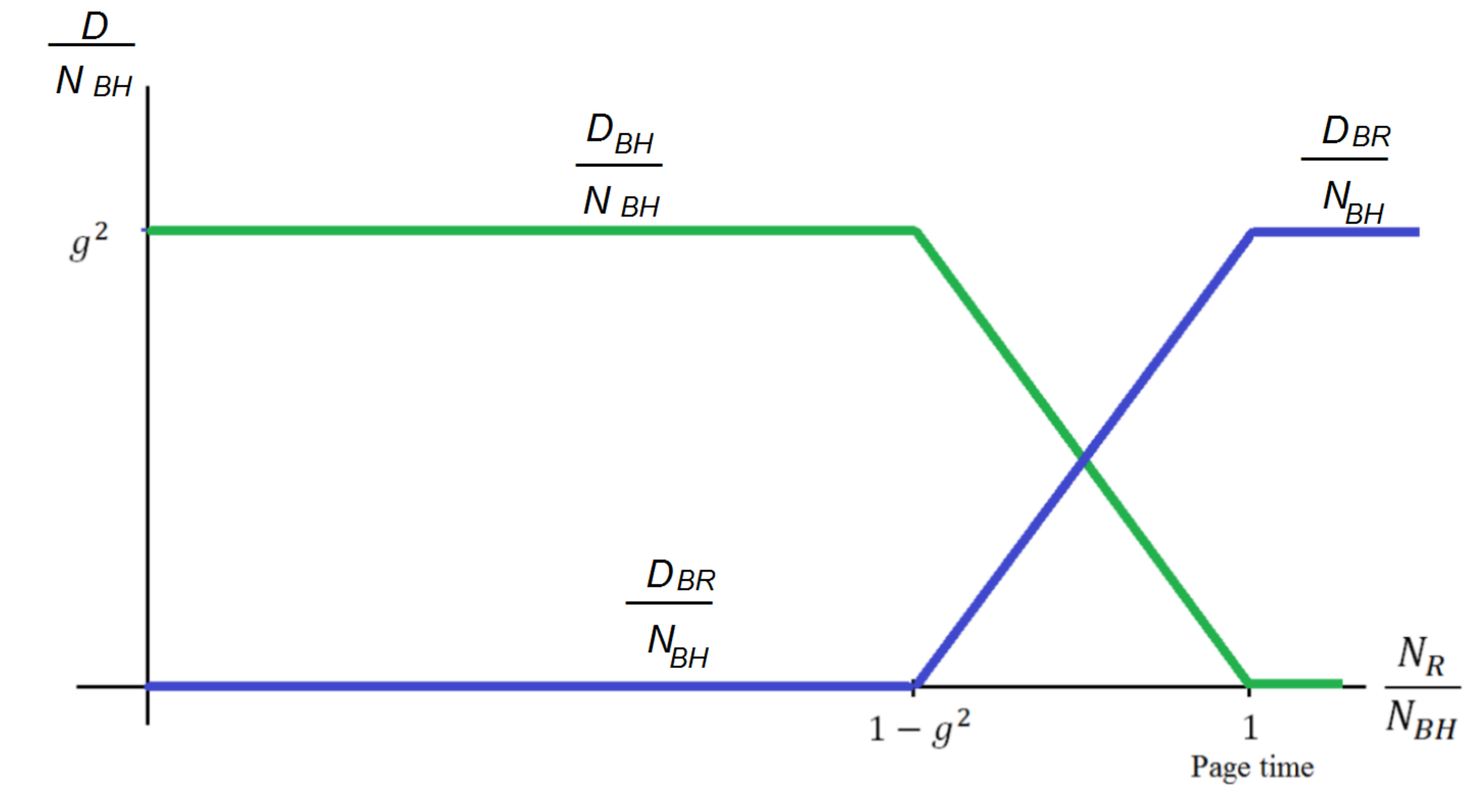}
\caption{  As the the \h\b \ distillable entanglement decreases, it is compensated by \r\b \ entanglement.        }
\label{4}
\end{center}
\end{figure}

 \r \ does provide a resource for entangled Bell pairs. Indeed, after the Page time the \dof \ $B_i$ continue to be entangled but with a subsystem of \r \ instead of \h. It is not hard to prove that the distillable entanglement between \b \ and the union  \h\r \ remains large enough to define partner modes for $B_i.$ In figure \ref{4}  the fractional distillable entanglement of \r\b \ is plotted alongside that of \h\b.
The total is in fact conserved. Before the cusp time the Bell pairs are shared between \h \ and \b. After the Page time they are shared between \r \ and \b. Between the cusp and Page times they are partly shared with \h \ and partly with \r, but the number of Bell pairs shared by \b \ is constant.

 After the Page time the degree of freedom that is maximally entangled with $B_i$ lives in \r \ and can be called $R_{B_i}.$ The hypothesis that at late times \a \ becomes redundant with the Hawking radiation would replace the    \h $\Longleftrightarrow$ \a \ mapping by an \r $\Longleftrightarrow$ \a \ mapping,
\be
 A_i = R_{B_i}.
\label{a=rb}
\ee
The relation between $B_i$ and  $R_{B_i}$ is very fine-grained and depends in detail on the precise initial state and dynamics of the black hole.

An identification  such as \ref{a=rb} would imply a radically greater localization-ambiguity than
$$
A = H_{B}.
$$
(and would also eliminate the need for firewalls).
Note however that  such large scale delocalization of information is already present in any holographic theory.

Now let's turn to the reasons that AMPS rejected \ref{a=rb}.

\subsection{Time Travel}
AMPS argued against \ref{a=rb} by invoking a thought experiment that leads to an apparent contradiction.
The experiment involves an observer Alice, who is equipped with a very powerful quantum computer (QC). The  input to the QC is the early half of the Hawking radiation. The output is a specific hidden qubit  that Alice can hold and manipulate.
It is assumed that Alice or her computer knows the exact initial state of the black hole, and the precise laws of evolution of all $N$ degrees of freedom comprising the system.

Furthermore, according to the AMPS argument, she can collect the radiation and use her quantum computer to distill $R_{B_i},$ which
by hypothesis is equal to $A_i.$ If $A_i$ refers to a field degree of freedom well after the Page time, then Alice knows information about $A_i$ long before $A_i$ has even happened. Alice then jumps into the black hole, carrying $R_{B_i},$ in time to meet the  original $A_i$ and its partner $B_i.$

The problem  is that Alice can check whether her version of $A_i$ (namely $R_{B_i}$) is entangled with $B_i.$ If it is then, by the monogamy of entanglement, $B_i$  cannot also be entangled with the original $A_i,$ and thus a firewall must exist. In fact there is no need for her to check since she is already sure the quantum computer has accurately distilled $R_{B_i}.$

Another way to express the paradox is that Alice's experiment is analogous to past time-travel. The degree of freedom $A_i$ has somehow appeared long in the past as $R_{B_i},$ and then traveled back to meet itself at $A_i.$ In this form the firewall is playing the role of Hawking's \it chronology protection agent. \rm

In \cite{Susskind:2012uw} an possible alternative chronology protection agent was remarked on; ``It may simply be physically impossible to distill $R_B$ out of the Hawking radiation in time to bring it back to meet $A$." The same idea had also been expressed by Bousso and by Harlow \footnote{The possibility that Alice's experiment could not be carried out in the required length of time came up repeatedly in conversations between Raphael Bousso, Daniel Harlow, myself, and several others, soon after the AMPS paper appeared. The feeling at that time was that  while difficult or impossible in practice, the principles of quantum
mechanics do not forbid the experiment.}.

The decoding part of the experiment is clearly impractical. However, unless it
violates a principle of physics such as locality, or the uncertainty principle, it must be allowed as a legitimate part of the AMPS argument. Therefore, to be consistent, the postulates of complementarity predict that the laws of physics  absolutely forbid the distillation of $R_{B_i}$  in less than the evaporation time.

\subsection{Strong Complementarity}

\begin{figure}[h!]
\begin{center}
\includegraphics[scale=.3]{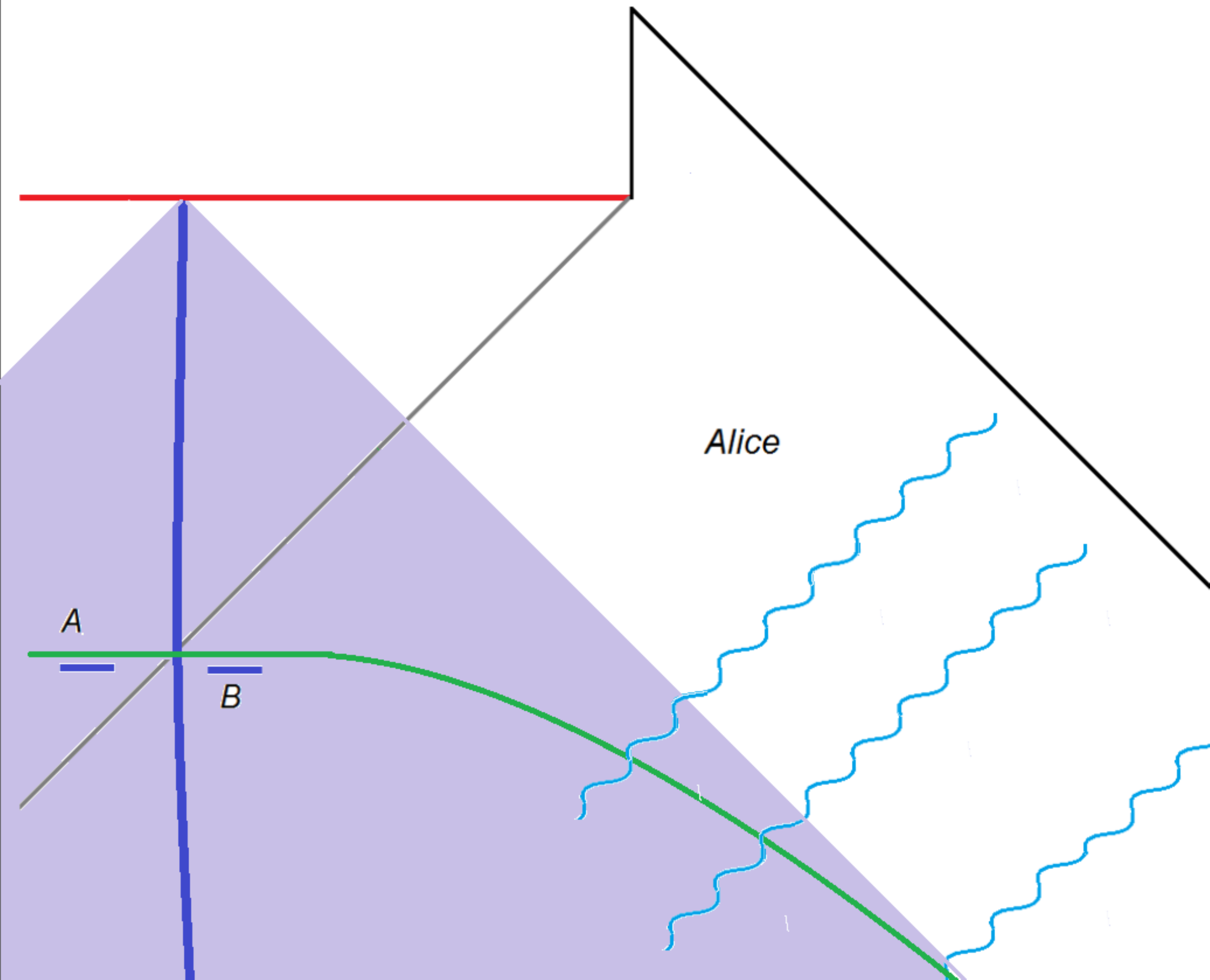}
\caption{  Alice's causal patch is shown as the light blue region. The green spacelike surface asymptotes to the light-like boundary of Alice's causal patch. It encompasses $A,$ $B,$ and the early half of the Hawking radiation.        }
\label{5}
\end{center}
\end{figure}

There is a formal version of the Alice's argument \cite{Bousso:2012as}. As Bousso and Harlow discuss, we can follow Alice to the end of her world-line  at point $\bf{p}$ on the singularity.  The causal past of  $\bf{p}$ defines Alice's causal patch
 shown in figure \ref{5}.

Her causal patch can be sliced by a family of space-like surfaces, one of  which passes through the modes $A,$  $B,$ and asymptotes to the light-like boundary of the Alice's causal patch. Complementarity requires entanglement between $B$ and $A$ in Alice's infalling frame. Assuming that $A$ and $B$ are well after the Page time, the space-like slice passing through them also
 intersects more than half the outgoing Hawking radiation.

\begin{figure}[h!]
\begin{center}
\includegraphics[scale=.3]{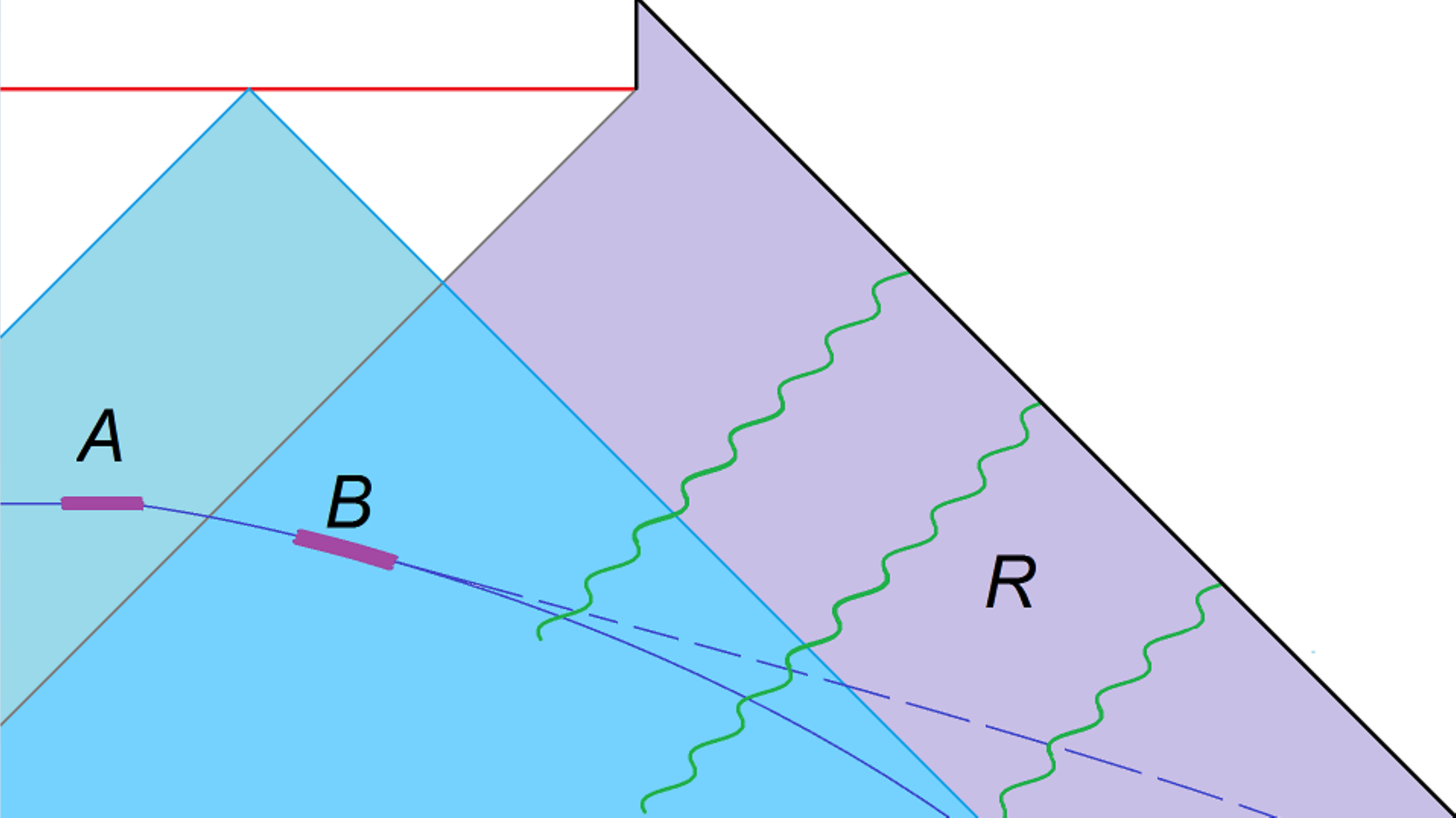}
\caption{  This figure is the same as the previous figure with Bob's causal patch superimposed. The dotted surface is a space-like hypersurface that asymptotes to light-like infinity. }
\label{AB}
\end{center}
\end{figure}

On the other hand the causal patch of Bob, who stays outside the black hole, contains $B,$ as well as the  outgoing radiation that was seen from Alice's patch (See figure  \ref{AB}), but it does not contain $A$. On Bob's space-like slice $B$ must be entangled with the outgoing radiation, i.e., with $R_B$

 The apparent contradiction is that  photons (of the first half of the radiation) which pass through Alice's surface, also pass through Bob's. Thus it would seem that if $B$ and $R_B$ are entangled on Bob's space-like slice, the same must be true on Alice's, and this leads to the unallowed polygamous entanglement that AMPS base their argument on. However, things may be more subtle than this.

\begin{figure}[h!]
\begin{center}
\includegraphics[scale=.3]{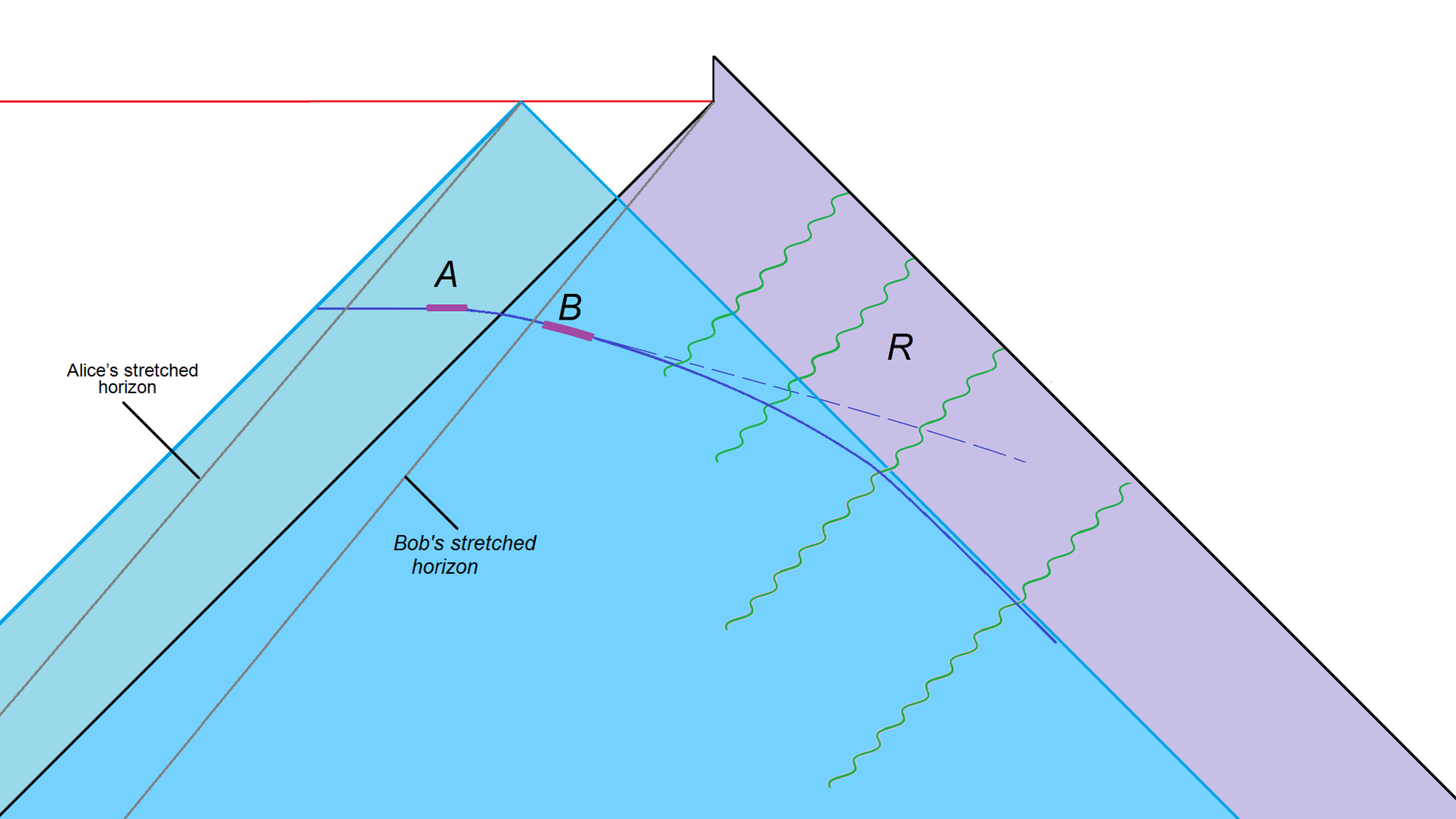}
\caption{  Both Alice and Bob have stretched horizons but they are not at the same place.      }
\label{ABSH}
\end{center}
\end{figure}

 Consider figure \ref{ABSH} which is the same as \ref{AB} except that stretched horizons have been drawn in.
 Bob's quantum description is from the outside of the black hole and the radiation region, the zone, and his stretched horizon. These \dof \ are coupled and Bob's  stretched horizon  is dynamically involved in the production of Hawking radiation. On the other hand, in Alice's infalling frame Bob's stretched horizon is absent; it is replaced by a displaced  stretched horizon that is shifted to the left. What is clear is that the two descriptions of the production of radiation cannot be exactly the same.

 At the level of coarse grained properties of the radiation, the descriptions  must match in the overlap region of the causal patches, out beyond the photon sphere where it is well understood that  infalling and stationary observers see the same photons. The question is whether the  two descriptions must match at the fine-grained level. In fact, Bousso and Harlow \cite{Bousso:2012as}  suggest that they do not. $B$ is a coarse-grained object and therefore Alice and Bob should agree about it, but  $R_B$ is extremely fine-grained. Since in Alice's frame there cannot be multiple entanglement, $R_B$ must not exist in her quantum description. In other words the radiation should not have the large-scale entanglements of a pure but scrambled state in her frame.

 In Bob's frame purity requires those entanglements, but Bob's description does not include $A.$

 But now it may be objected that we have an observable contradiction: Bob studies the photons that he sees and concludes that there is $(B,  R_B)$ entanglement. Alice studies the same photons and says there is not $(B,  R_B)$ entanglement.

Bousso  and Harlow \cite{Bousso:2012as} have argued against this conclusion, advocating a strong form of complementarity that can be described by saying each causal patch has its own quantum description. In Alice's quantum mechanics $B$ is entangled with $A$ and not with the outgoing radiation. In Bob's description $B$ is entangled with $R_B.$

In fact we will see that the Harlow-Hayden conjecture protects against any real contradiction and allows the Bousso-Harlow strong complementarity to be consistent.

\subsection{The Harlow-Hayden Conjecture}

Hayden and Harlow (HH) have presented evidence suggesting that the Alice experiment does, in fact,  violate fundamental physical principles \cite{HH}. They conjectured that the  distillation of $R_{B_i}$  in less than exponential time (exponential in $N$) is impossible with a QC that satisfies the principles of quantum mechanics and relativity. Satisfying quantum mechanics is, of course, part of being a quantum computer, but special and general relativity also impose limitations such as locality and  holographic limitations on information.

\begin{figure}[h!]
\begin{center}
\includegraphics[scale=.3]{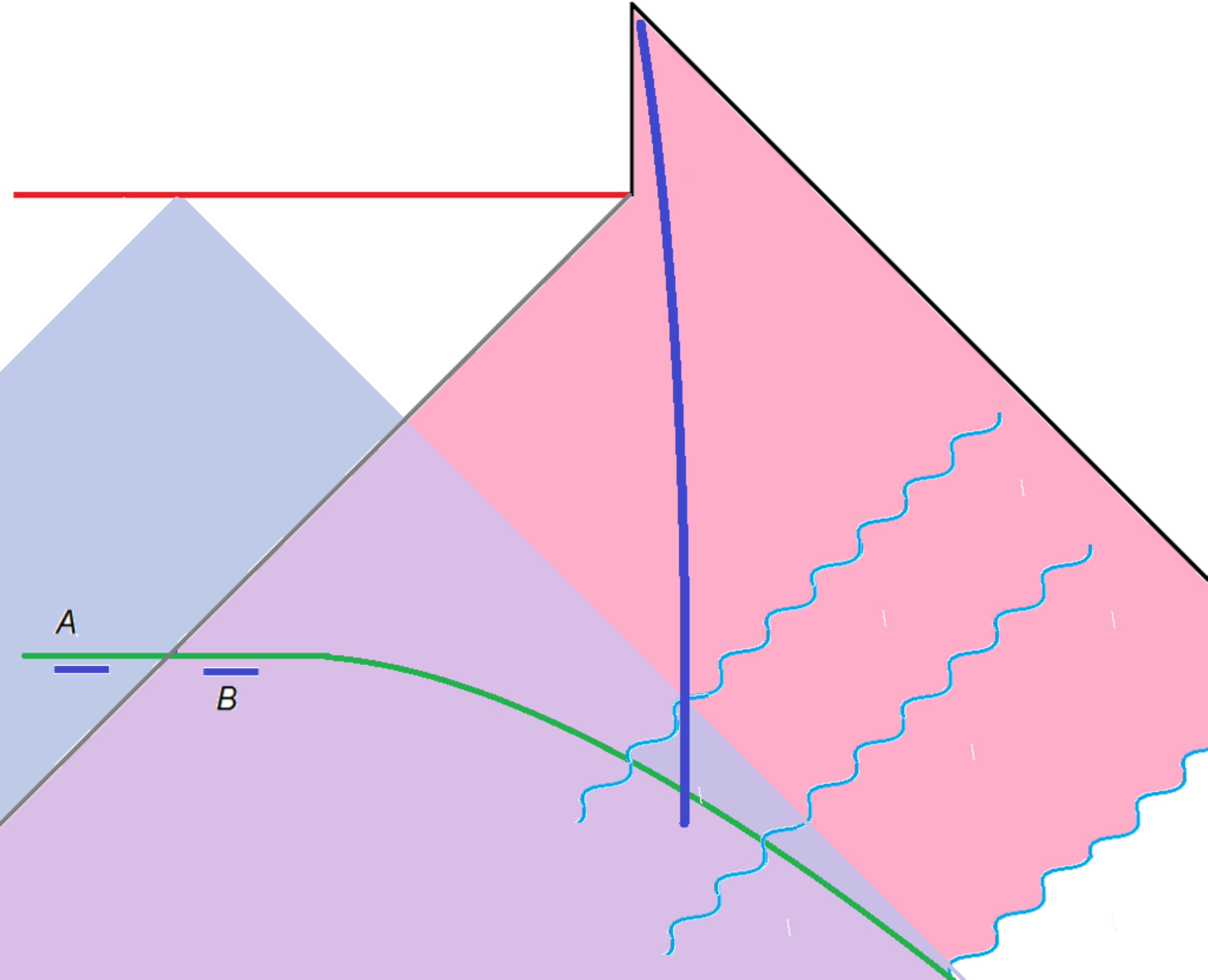}
\caption{ If the HH conjecture is true, $R_B$ cannot be distilled until the black hole has evaporated. The distillation can only be done by the time Alice arrives in the uppermost corner of the Penrose diagram. }
\label{6}
\end{center}
\end{figure}

If the HH conjecture is correct, then the operational limitation on Alice's experiment are shown in
Figure \ref{6}.
If the extraction of $R_{B_i}$ takes exponential time, then an observer  can only distill $R_{B_i}$  after she has reached the corner in the diagram where the black hole has already evaporated. In other words whatever mechanism implements the HH conjecture serves as a chronology protection agent.

 The Harlow Hayden argument is technical \cite{HH} but let me discuss the issues.
Here is the problem Alice faces:
To begin with, the system of $N$ qubits starts in a known pure state, schematically representing the initial state of the black hole. As an example the initial state could be $|\Psi_0\ra = |00000000.....\rangle.$ Subsequently the internal dynamics of the black hole, represented by a unitary scrambling operator $U,$ scrambles the state,
\be
|\Psi \ra = U |\Psi_0 \ra
\ee
The matrix $U$ is drawn from a statistical ensemble which is not strictly random. Scrambling does not require the state $|\Psi \ra $ to be typical, but only that the density matrices of small subsystems be random (proportional to the identity to high accuracy). In the quantum-information jargon, it is sufficient that $U$ be a unitary 2-design.

The unitary 2-design ensemble (U2) is  very far from the Haar-random ensemble that would ordinarily define random matrices. To see how different they are, we can compare the number of 2-body interactions that  must take place to implement U2, and the number needed to Haar-randomize. The answer is that the lower bound  to obtain U2 randomness is $N \log N$ interactions (the bound is achievable), while Haar randomness requires an exponential number of interactions \cite{Hayden:2007cs}.
However, although U2 and Haar-randomness are very different, they lead to the same statistical properties for small subsystems, and for that reason U2 matrices scramble, and they can do so in the scrambling time of order $N \log{N}.$

Once the system is scrambled, evaporation separates $(N-M)$ qubits into \r. The case of interest is when $(N-M)$ is of order $N$ and somewhat larger than $N/2.$  The remaining black hole ($ \cal{H} \cup  \cal{B}) $) is represented by $M$ qubits.

Select one qubit from the remaining black hole and let it represent $B_i.$ To be precise, the symbol $B_i$ represents the set of operators that act on the qubit in the same sense that the symbol $\sigma$ represents the three Pauli spin operators acting on a spin. The partner qubit $R_{B_i}$ is hidden among the $(N-M)$ qubits in \r, and Alice's job is to isolate or distill it, and convert it to a qubit that she can hold and manipulate. To do that, she can make use of her QC to physically implement an unscrambling unitary $V$ operator that acts on the subsystem of $(N-M)$ qubits. The purpose of $V$ is to act on $R_{B_i}$ and convert it to a particular qubit, for example the last qubit, $q_0$ which Alice is waiting to grab. Thus $V$ is defined by
\be
V^{\dag} R_B V = q_0
\ee
The operator $V$ is of course not unique but this is not too important.

Alice does not have to search through all unitary operators $V$ until she finds one that accomplishes the task.  Knowing the initial state, and the scrambling matrix $U,$ she can calculate $V.$  In fact she could have done this long before the black hole  formed. She has all of infinite past time to prepare for the experiment.
Alice's limitation is not the difficulty of determining $V, $ but of actually using her QC to implement $V,$ in other word to physically apply it to the system of $M$ qubits.

The original scrambling operator $U$ is drawn from the U2 ensemble for $N$ qubits, and only takes polynomial time to implement. If Alice had all $N$ qubits to work with she could unscramble the system with the inverse matrix $U^{\dag}$ which can be implemented in polynomial time.
 Because Alice has only $(N-M)$ qubits to work with, the matrix $V$ is much more generic than a unitary 2-design, and more difficult to implement. This is analogous to the classical problem of time-reversing a complex chaotic system if even  a small amount of information is lost before reversing it.

 If $V$ is sufficiently generic it will take an exponential number of gates (exponential in $(N-M)$) to implement $V.$ Parallel computing does not help much.
HH argue from experience with similar problems in quantum information theory,  that the number of gates that have to operate in order to implement $V$ is indeed exponentially  in $(N-M).$  Therefore, they conjecture that it takes exponential time for Alice to distill $R_{B_i}.$  (The exponential time is analogous to the recurrence time for a classical system.) \\

Given that the internal dynamics of a black hole can scramble in a short time, why then does it take so long to de-scramble? There is a close classical analog from the  dynamics of complex  chaotic systems. Consider a gas of $N$ molecules in a sealed box. $(N-1)$ of the molecules are identical of type $a,$ and one is different of type $b$. Now start the system in some very non-generic state; for example all $(N-1)$ $a$-type molecules at rest in one corner of the box, and the single $b$-type molecule at some other location $x_b$ with a very large energy. Let the system evolve for a (polynomial) time until it appears  to be in thermal equilibrium.
The goal of the exercise is for Alice to recover $x_b.$

Alice is equipped with an incredibly powerful  computer, which can if necessary integrate the equations backward, but her eyesight is somewhat blurry. The input to the computer is the the final particles at some time after the system looks to be in equilibrium. The question is how long must it take the computer to recover the initial location of $b$?

The answer is---no longer than it took to come to equilibrium. If the computer is powerful enough it can simply reverse the process and bring the system back to its initial state. From that configuration, which is after all, special and easily recognizable, Alice just reads off $x_b.$ By assumption that can be done in polynomial time.

One might ask why can't Alice look carefully at the system at the final time, and recognize what the initial state was? She can't because she is a bit myopic; the phase point is a bit blurry; and any small error will, if run backward, exponentially grow. But if the system is run all the ways back, the initial state is sufficiently distinct that Alice's vision is good enough, and she can recover $x_b$ in polynomial time.

But now lets suppose that
 just before the system is run backward, we  change the hamiltonian by dynamically decoupling  $M$ $a$-type particles.  The computer must work with the remaining $(N-M)$ particles. Then running backward will  not result in a distinctive configuration that allows Alice to  read off the information we are looking for in polynomial time. Instead the system will stay in  scrambled equilibrium for a recurrence time of order $ \exp(N).        $  Only after such a long time will the
system  fluctuate to a highly non-generic state, which Alice can read.

 One might argue that the limitations of Alice's eyesight are irrelevant; she can just buy a better pair of eyeglasses. That's where quantum mechanics comes in. The uncertainty principle means that the fuzzyness is fundamental.

 In the real case of  decoding $R_B,$ if Alice had all the degrees of freedom of \h, \b, and \r, she could run $U$ backward in polynomial time (less than the evaporation time). But because she only gets to look at the qubits in \r \
 her job is much harder, and the guess is that it takes exponential time to distill $R_B.$

Considerations of this type led  Harlow and Hayden to the following conjecture:

\bigskip

\it The minimum  time that it takes to distill  qubit $R_{B_i},$  is exponentially grows with $N.$ \rm

\bigskip

At the time of writing this talk, the Harlow-Hayden conjecture is still a conjecture. However, if  true, it implies a fundamental physical constraint---call it computational complexity---that would prevent Alice from carrying out her experiment.  For an ordinary \S \ black hole the total time before it evaporates is of only of order $N^3.$ Thus the truth of the HH conjecture would undermine the thought  experiment designed to prove the existence of firewalls.

A possible way out of the computational complexity constraint was suggested in \cite{Susskind:2012uw} and probably by several other people. Alice may try to slow down the evaporation process while her quantum computer is distilling $R_{B_i}.$ One way to do that would be to surround the black hole by mirrors to keep it from radiating.
But as Harlow and Hayden argue, this might not help Alice if the decoding time is exponential. An exponential time scale has multiple meanings for a complex closed system. For one thing the time scale for resolving tiny energy differences between neighboring states is of order $\Delta t =\frac{1}{\Delta E} .$ For a system of entropy $S$ this is  equal to $e^S \sim e^N.$  Of greater relevance, over such time scales Poincare recurrences will repeatedly occur, undoing and re-collapsing the black hole. It is unlikely that the identity of a mode $A_i$ has any meaning over such long times.

Now let's return to Alice and Bob's Disagreement: Alice says no $B, \ R_B$ entanglement in her frame. Bob says that there is $B, \ R_B$ entanglement in his frame. That's of course the nature of complementarity, but can their disagreement lead to an observational conflict? The answer is no if the HH conjecture is correct; the difference between states with massive entanglement like $|\Psi \ra$ and those with none like $\rho$ can only be detected long after it is too late for Alice and Bob to communicate.

\section{Conclusion}
The AMPS paradox is currently forcing a rethinking of how, and where, information is stored in quantum gravity. The possible answers range from more or less conventional localization (proximity and  firewalls) to the radical delocalization of $A=R_B.$

 The argument in favor of the proximity postulate  assumes the possibility of an ``Alice experiment", which in some respects resembles time travel to the past. From this perspective, the firewall would function as a chronology protection agent, but at the cost of the destruction of the interior of the black hole.
The Harlow-Hayden conjecture opens an entirely new perspective on chronology protection. It is based on the extreme fine-grained character of  information that Alice needs to distill before ``returning to the present."

Fine-grained information is something that has never been of much use in the past, given how hard it is to extract. But there is clearly a whole world of fine-grained data stored in the massive entanglements of  scrambled pure states. That world is normally inaccessible to us, but if BHC is correct, it is  accessible to an observer who passes through the horizon of a black hole. Thus complementarity implies a duality: \\

\it Ordinary coarse-grained information in an infalling  frame is dual to the fine-grained information in the exterior description. \rm   \\

Or, as expressed earlier: \\

\it Ground-state entanglements in the  infalling Minkowski vacuum, are dual to
scrambled entanglements of the exterior thermal description. \rm \\

The HH conjecture also  represent a new principle, based on quantum-mechanical computational complexity: \\

 \it Extracting fine-grained information cannot be done in less than exponential time, comparable to the time scale for Poincare recurrences. Therefore it cannot be communicated back  to the black hole before it evaporates.\rm \\

It's obviously premature to declare the paradox resolved, but the validity of the HH conjecture
  would allow the strong complementarity of Bousso and Harlow \cite{Bousso:2012as} to be consistent, without the need for firewalls. For these reasons I believe that   BHC, as originally envisioned by Preskill, 't Hooft, and Susskind-Thorlacius-Uglum,
is still alive and kicking.

\section*{Acknowledgements}

 Daniel Harlow, and Patrick Hayden, and Raphael Bousso have been very generous in sharing their ideas concerning the limits on
distilling $R_B$ and the idea of strong complementarity as well as the $A=R_B$ hypothesis and its potential pitfalls.
The ideas I've expressed are largely due to them.

I also owe a great deal to Douglas Stanford for help in understanding  distillable entanglement and  the HH conjecture.

\end{document}